\documentclass[twocolumn,epjc3]{svjour3}          

\RequirePackage[T1]{fontenc}
\smartqed  

\RequirePackage{graphicx}
\RequirePackage{epstopdf}
\RequirePackage[colorlinks,citecolor=blue,urlcolor=blue,linkcolor=blue]{hyperref}
\RequirePackage{amssymb}
\RequirePackage{slashed}
\RequirePackage{amsmath}
\RequirePackage{subfigure}
\hypersetup{pdfstartview=}

\makeatother

\begin{document}

\title{Contributions to $ZZV^\ast$ ($V=\gamma,Z,Z'$) couplings from $CP$ violating flavor changing couplings}
\author{A. I. Hern\'andez-Ju\'arez    \thanksref{e1,addr1}\and
        A. Moyotl  \thanksref{e2,addr2}\and
        G. Tavares-Velasco  \thanksref{e3,addr1}
}

\thankstext{e1}{alaban7\_3@hotmail.com}
\thankstext{e2}{agustin.moyotl@uppuebla.edu.mx}
\thankstext{e3}{gtv@fcfm.buap.mx}

\institute{Facultad de Ciencias F\'isico-Matem\'aticas,\\
  Benem\'erita Universidad Aut\'onoma de Puebla,\\
 C.P. 72570, Puebla, Pue., Mexico \label{addr1} \and Ingenier\'ia en Mecatr\'onica,\\ Universidad Polit\'ecnica de Puebla,\\ Tercer Carril del Ejido Serrano s/n, San Mateo Cuanal\'a, Juan C. Bonilla,\\ Puebla, Puebla, M\'exico \label{addr2}}

\date{Received: date / Accepted: date}
\maketitle
\begin{abstract}
The one-loop  contributions to the trilinear neutral gauge boson couplings $ZZV^\ast$ ($V=\gamma,Z,Z'$),  parametrized in terms of one $CP$-conserving $f_5^{V}$ and one $CP$-violating $f_4^{V}$ form factors,  are calculated in models with  $CP$-violating  flavor changing neutral current couplings  mediated by the $Z$ gauge boson and an extra neutral gauge boson $Z'$. Analytical results are presented in terms of  Passarino-Veltman scalar functions.
Constraints  on the vector and axial couplings of the $Z$ gauge boson $\left|g_{{VZ}}^{tu}\right|< 0.0096$ and $\left|g_{{VZ}}^{tc}\right|<0.011$ are obtained from the current experimental data on the $t\rightarrow Z q$ decays.
It is found that in the case of the $ZZ\gamma^\ast$ vertex the only non-vanishing form factor is $f_5^{\gamma}$,
which can be of the order of $10^{-3}$, whereas for the $ZZZ^\ast$ vertex both  form factors $f_5^{Z}$ and   $f_4^{Z}$  are non-vanishing and can be of the order of $10^{-6}$ and $10^{-5}$, respectively. Our estimates for $f_5^{\gamma}$ and $f_5^{Z}$  are smaller than those predicted by the standard model, where  $f_4^{Z}$  is absent up to  the one loop level. We also estimate the $ZZ{Z'}^{*}$ form factors arising from  both  diagonal  and non-diagonal $Z'$ couplings within a few extension models. It is found that in the diagonal case $f_{5}^{Z'}$ is the only non-vanishing form factor and  its real and imaginary parts can be of the order of   $10^{-1}-10^{-2}$ and $ 10^{-2}-10^{-3}$, respectively, with the dominant contributions arising from the light quarks and leptons. In the non-diagonal  case   $f_{5}^{Z^\prime}$ can be of the order of $10^{-4}$, whereas  $f_4^{Z'}$ can reach values as large as $10^{-7}-10^{-8}$, with the largest contributions arising from the $Z'tq$ couplings.

\keywords{Trilinear  gauge boson couplings \and CP violation  \and Flavor changing neutral currents }

\end{abstract}

\section{Introduction}
\label{Intro}
Trilinear  gauge boson couplings (TGBCs)  have long been the subject of considerable interest both theoretically and experimentally. In the experimental area, constraints on the corresponding form factors were first obtained at the LEP \cite{Acciarri:1998iw,Abbiendi:2000cu,Abdallah:2007ae} and the Tevatron 
\cite{Abe:1994fx,Abachi:1997xe,Aaltonen:2011zc} colliders, whereas the current bounds were
extracted from the LHC data at  8 TeV  \cite{CMS:2014xja,Khachatryan:2015pba,Aaboud:2016urj} and  13 TeV  \cite{Aaboud:2017rwm,Sirunyan:2017zjc,Sirunyan:2020pub} by the ATLAS and CMS collaborations. Among TGBCs are of special interest the ones involving only  neutral gauge bosons, namely, the trilinear neutral gauge boson couplings   (TNGBCs) $ZZV^\ast$ and $Z\gamma V^\ast$ ($V=Z$,$\gamma$), which can only arise up to the one-loop level in renormalizable theories and  have been widely studied within the standard model (SM) and beyond.   The SM contributions to TNGBCs were studied in Refs. \cite{Gounaris:2000tb,Choudhury:2000bw}, whereas new physics contributions have been studied within several extension models, such as the minimal supersymmetric standard model (MSSM)\cite{Gounaris:2000tb,Choudhury:2000bw,Gounaris:2005pq}, the $CP$-violating two-Higgs doublet model (2HDM) \cite{Corbett:2017ecn,Belusca-Maito:2017iob,Grzadkowski:2016lpv}, models with axial and vector fermion couplings \cite{Corbett:2017ecn}, models with extended scalar sectors \cite{Moyotl:2015bia}, and also via the effective Lagrangian approach \cite{Larios:2000ni}. In the theoretical side, the phenomenology of TNGBCs at particle colliders was widely studied long ago 
\cite{Gounaris:1999kf,Baur:1992cd,Alcaraz:2001nv,Walsh:2002gm,Atag:2004cn,MoortgatPick:2005cw,Gounaris:2005pq,GutierrezRodriguez:2008tb} and also has been of interest lately \cite{Ananthanarayan:2011fr,Senol:2013ym,Ananthanarayan:2014sea,Ellis:2019zex}.  Even more,  study of the potential effects of TNGBCs at future colliders has been the source of renewed interest very recently 
\cite{Rahaman:2017qql,Rahaman:2016pqj,Behera:2018ryv,Rahaman:2018ujg,Yilmaz:2021qnv}.  TNGBCs, which require  one off-shell gauge boson at least to be non-vanishing due to Bose statistics, are  induced through dimension-six and dimension-eight operators \cite{Gounaris:2000tb,Larios:2000ni,Gounaris:2000dn,Degrande:2013kka} and can be parametrized in a model independent way by two $CP$-even and two $CP$-odd form factors. In the SM, only the $CP$-conserving form factors arise at the one-loop level of perturbation theory, whereas the $CP$-violating  ones are absent at this order and  require new sources of $CP$ violation \cite{Gounaris:2000tb,Gounaris:2000dn}. In the SM, $CP$ violation is generated via the Cabbibo-Kobayashi-Maskawa (CKM) mixing matrix, though the respective amount is not enough to explain the asymmetry between matter and anti-matter in the universe, {\it i.e.} the so-called baryogenesis problem. Therefore,  new sources of $CP$ violation are required, which is in fact one of the three Sakharov's conditions to explain the  baryon asymmetry of the universe \cite{Sakharov:1967dj}.  In this work we are interested in the study of possible $CP$-violating effects in the TNGBCs via tree-level  flavor changing neutral currents (FCNCs) mediated by the $Z$ gauge boson \cite{Moyotl:2017ljz}, which are forbidden in the SM but can arise in several SM extensions \cite{Buchalla:2000sk,Mohanta:2005gm}.  The possible effects of $Z$-mediated FCNC couplings on $CP$-conserving TNGBCs have already been studied \cite{Gounaris:2000tb}, nevertheless, possible contributions to the $CP$-violating ones have not been reported yet to our knowledge.  These new contributions are worth studying as they could shed some light in the path to a more comprehensive SM extension.

Possible evidences of  new heavy gauge boson have been searched for at the LHC  by the CMS collaboration \cite{Sirunyan:2019jbg}, which has been useful to set bounds on the masses of new neutral and charged heavy vector bosons. Such particles are predicted by a plethora of SM extensions with extended gauge sector, for instance,  little Higgs models \cite{Perelstein:2005ka,delAguila:2011wk}, 331 models \cite{Dong:2014wsa}, left-right symmetric models \cite{Mohapatra:1974gc}, etc. Some of these models allow tree-level FCNCs mediated by a new neutral gauge boson, denoted from now on by $Z^{\prime}$ \cite{Arhrib:2006sg}, which means that $CP$-violating contributions to  $VZZ'^\ast$ couplings ($V=\gamma$, $Z$) are  possible. To our knowledge TNGBCs with new neutral bosons have not received much attention in the literature up to now, though decays of the kind $Z^{\prime}\rightarrow VZ$ ($V=\gamma$,$Z$) \cite{CortesMaldonado:2011pi} and $Z^{\prime}\rightarrow \gamma A_H$ \cite{CortesMaldonado:2011dv}  were already studied. Here $A_H$ stands for a heavy photon.

In this work we present a study of  the one-loop contributions to the most general TNGBCs $ZZV^\ast$  ($V=\gamma$, $Z$, $Z^{\prime}$) arising from a generic model allowing tree-level FCNCs mediated by the SM $Z$ gauge boson and a new heavy neutral gauge boson $Z'$. The rest of this presentation is as follows. In Sec. II we present a short review of the analytical structure of TNGBCs along with  the theoretical framework of the FCNCs $Z$ and $Z'$ couplings via a model independent approach.  Section III is devoted to the calculation of the one-loop contributions to the $CP$-conserving and $CP$-violating $ZZV^\ast$  ($V=\gamma$, $Z$, $Z^{\prime}$) couplings, for which we use the Passarino-Veltman reduction scheme. In Sec. IV we present the numerical analysis and discussion, whereas the conclusions and outlook are presented in Sec. V.

\section{Theoretical framework}
\subsection{Trilinear neutral gauge boson couplings}
\begin{figure}[!htb]
  \centering
\includegraphics[width=8cm]{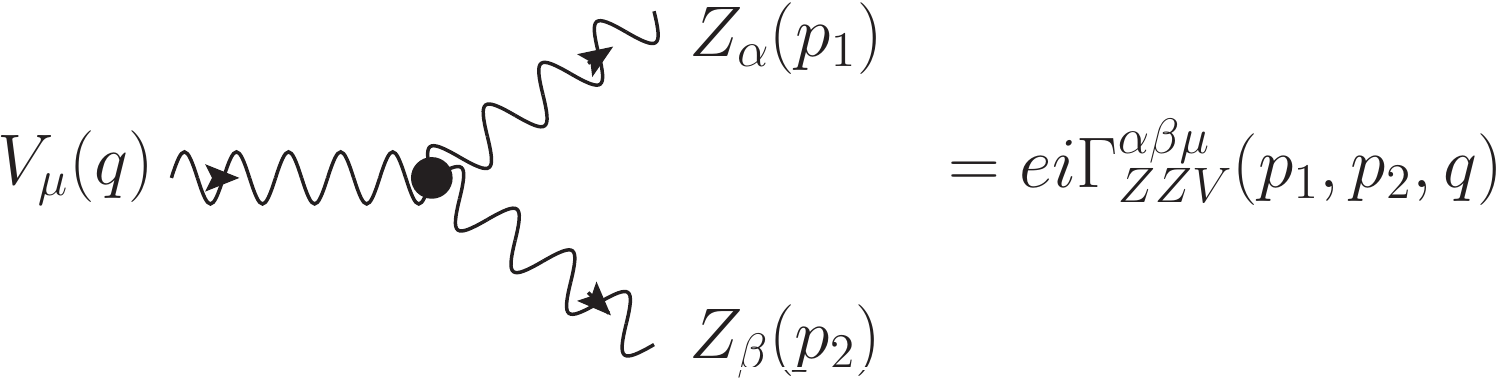}
  \caption{Nomenclature for the TNGBCs $ZZV^*$ ($V=\gamma$, $Z$, $Z'$). }\label{FeynmanDiagram1}
\end{figure}

We now turn to discuss the Lorentz structure of TNGBCs, which are induced by dimension-six and dimension-eight operators. In this work we only focus on the contribution of dimension-six operators as it is expected to be the dominant one. In particular, the TNGBC  $ZZV^*$ ($V=\gamma$, $Z$) coupling  can be parametrized by two form factors:
\begin{align}
\label{vertexZZV}
  \Gamma_{ZZV^\ast}^{\alpha\beta\mu}\left(p_1,p_2,q\right)&= \frac{i (q^2-m_V^2)}{m_Z^2}\Big[f_4^V\left(q^\alpha g^{\mu\beta}+q^\beta g^{\mu\alpha}\right)\nonumber \\ 
  &-f_5^V \epsilon^{\mu\alpha\beta\rho}\left( p_1 -p_2\right)_{\rho} \Big],
\end{align}
where we have followed  Ref. \cite{Gounaris:2000tb}, with the notation for the gauge boson four-momenta being depicted in Fig. \ref{FeynmanDiagram1}. From Eq. \eqref{vertexZZV} it is evident that when the $V^*$ gauge boson becomes on-shell ($q^2=m_V^2$), $\Gamma_{ZZV^\ast}^{\alpha\beta\mu}\left(p_1,p_2,q\right)$ vanishes, which is due to Bose statistics and angular momentum conservation. The general form of this vertex for three off-shell gauge bosons  can be found in   \cite{Gounaris:2000dn,Larios:2000ni}.  The form factor $f_5^V$ is $CP$-conserving, whereas $f_4^V$ is $CP$-violating.  The former is the only one  induced at the one-loop level in the SM via a fermion loop since $W^\pm$ boson loops give vanishing contributions \cite{Gounaris:2000tb}. It was found that $f_5^V$  decreases quickly as $q^2$ becomes large \cite{Gounaris:2000tb}. The current bounds on the form factors $f_4^V$ and $f_5^V$ ($V=Z$, $\gamma$) were obtained by the CMS collaboration at $\sqrt{s}=13$ TeV \cite{Sirunyan:2020pub}:

\begin{align}
    -0.00066< &f_4^Z<0.0006,  \\ -0.00055<&f_5^Z<0.00075,    \\
    -0.00078< &f_4^\gamma<0.00071,\\   -0.00068<&f_5^\gamma<0.00075.
\end{align}

These constrains are of the order of the SM prediction for the $CP$-conserving form factors \cite{Gounaris:2000tb}. Thus, corrections to the form factors $f_5^V$ and $f_4^V$ ($V=\gamma$, $Z$) from models of new physics might play an important role.

 As far as TNGBCs with new gauge bosons are concerned, they remain almost unexplored. For our purpose, following Eq. \eqref{vertexZZV}, we will parametrize the $ZZ{Z^{\prime}}^\ast$ coupling as follows

\begin{align}
\label{ZZZHvertex}
  \Gamma_{ZZ{Z^{\prime}}^\ast}^{\alpha\beta\mu}\left(p_1,p_2,q\right)&= \frac{i q^2}{m_{Z^{\prime}}^2}\Big[f_4^{Z'}\left(q^\alpha g^{\mu\beta}+q^\beta g^{\mu\alpha}\right) \nonumber \\
  &-f_5^{Z'} \epsilon^{\mu\alpha\beta\rho}\left( p_1 -p_2\right)_{\rho} \Big],
\end{align}
where we only consider the contributions of the dimension-six operators given in Ref. \cite{Gounaris:2000dn} and  have replaced the electromagnetic $F_{\mu\nu}$ tensor by $Z'_{\mu\nu}=\partial_\mu Z'_{\nu} -\partial_\nu Z'_{\mu}$ in the operator basis that induce the $ZZ\gamma^\ast$ vertex. We also set the  energy scale that corrects the operator dimension to the new physics scale $m_{Z'}$.  Of course, the form factor $f_4^{Z'}$ ($f_5^{Z'}$) is  $CP$ violating ($CP$-conserving). We also note in Eq. \eqref{ZZZHvertex} that this TNGBC does not vanish  for an on-shell $Z'$ gauge boson. In fact,  the $Z'$ gauge boson can decay into  a  $Z$ gauge boson pair if kinematically allowed. We will see below that our calculation is consistent with the Lorentz structure presented in Eq. \eqref{ZZZHvertex}.

\subsection{ FCNCs mediated by the $Z$ and $Z'$ gauge bosons}

Beyond the SM, there are some extension theories that allow FCNC couplings mediated by the $Z$ gauge boson  \cite{Buchalla:2000sk,Mohanta:2005gm}. Such an interaction can be expressed by the following  Lagrangian

\begin{align}
\label{LagrangianZ}
    \mathcal{L}&=-\frac{e}{2 s_W c_W}Z^\mu\overline{F}_{i}\gamma_\mu\left( {g^{i}_{VZ}}-\gamma^5g^i_{AZ} \right)F_i \nonumber\\
   &   -\frac{e}{2 s_W c_W}Z^\mu\overline{F}_{i}\gamma_\mu\left( {g^{ij}_{VZ}}-\gamma^5{g^{ij}_{AZ}} \right)F_j  {,}
\end{align}
where $F_{i,j}$ are SM fermions in the mass eigenbasis. Here $g^{i}_{VZ,AZ}$  are the diagonal SM couplings, whereas  the non-diagonal couplings  $g^{ij}_{VZ,AZ}$ $(i\ne j)$   will be taken as complex since we are interested in the $CP$-violating contribution. The latter must fulfil  $g^{ij\ast}_{VZ,AZ}=g^{ji}_{VZ,AZ}$ because of their hermiticity. It is also customary to express the Lagrangian of Eq. \eqref{LagrangianZ} in terms of the left- and right-handed projectors $P_L$ and $P_R$, with the chiral couplings denoted by $\epsilon^Z_{L_{ij}, R_{ij}}$, which are given in terms of the vector and vector-axial couplings $g^{ij}_{VZ,AZ}$ as follows

\begin{equation}
\label{chiral}
g^{ij}_{VZ,AZ}=   \frac{\epsilon^Z_{L_{ij}}\pm\epsilon^Z_{R_{ij}}}{2}.
\end{equation}
Below we will use both the $g^{ij}_{VZ,AZ}$ and $\epsilon^Z_{L_{ij},R_{ij}}$   parametrizations. The former is useful for the purpose of comparison with previous works, whereas the latter is best suited for our numerical analysis.

Possible phenomenological implications of FCNC couplings mediated by the $Z$ gauge boson have been studied within the SM \cite{Clements:1982mk,Hou:1986ur,Bernabeu:1987me,AguilarSaavedra:2004wm}, fourth-generation models \cite{Silverman:1991fi,Clements:1982mk,Hou:1986ur,Alok:2012xm,Mohanta:2010yj}, See-Saw models \cite{Bernabeu:1993up}, etc. Such FCNC couplings have been constrained via the   $b\rightarrow s$ transition \cite{Buchalla:2000sk,Mohanta:2005gm}, Kaon decays \cite{Buras:1998ed,Colangelo:1998pm}, $B-\overline{B}$ mixing \cite{Silverman:1991fi}, and $B^0$ decays \cite{Nir:1990yq,Giri:2003jj}. More, recently  constraints on  FCNC top quark decays $t\rightarrow q Z$ were reported by the ATLAS Collaboration at $\sqrt{s}=13$ TeV \cite{Aaboud:2018nyl}.

As for  models with FCNC mediated by a new neutral gauge boson, which we generically have denoted by  $Z'$, they have been widely studied in the literature \cite{Langacker:1991pg,Cakir:2010rs}. In Table \ref{FCNCmodels} we present a summary of some of the more popular models that predict a new $Z'$ gauge boson.

\begin{table*}[!htb]
\centering
\caption{Models in which there is a new neutral gauge boson with FCNC couplings \cite{Langacker:1991pg}.\label{FCNCmodels}}
\begin{tabular}{ccc}
\hline\hline
New heavy neutral gauge boson&Model&Gauge group\\
\hline\hline
$Z_h$&Sequential $Z$&$SU_L(2)\times U_Y(1)\times U'(1)$\\
$Z_{LR}$&Left-right symmetric&$SU_L(2)\times SU_R(2)\times U_Y(1)$\\
$Z_\chi$&Gran Unification&$S0(10)\rightarrow SU(5)\times U(1)$\\
$Z_\psi$&Superstring-inspired&$E_6\rightarrow SO(10)\times U(1)$\\
$Z_\eta\equiv \sqrt{3/8} Z_\chi-\sqrt{5/8 }Z_\psi$&Superstring-inspired& $E_6\to$ Rank-5 group\\
\hline\hline
\end{tabular}
\end{table*}

To describe the   FCNC $Z'$ couplings we follow the formalism presented in \cite{Arhrib:2006sg,Langacker:1991pg} and introduce an effective Lagrangian analogue to that of Eq. \eqref{LagrangianZ} in terms of the chiral couplings $\epsilon^{Z'}_{L_{ij}, R_{ij}}$,
where  $i$ and $j$ now run over all the SM fermions $f_{SM}=\nu_i,\ell_i,u_i,d_i$. though there can  also be new hypothetical fermions. We thus write
\begin{equation}
\label{LagrangianZ'}
\mathcal{L}^{FCNC}_{Z'}=-g_{Z'} \sum_{i=f_{SM}}
Z'_\mu \bar{\mathbf F}_i \gamma^\mu  \left( {\boldsymbol\epsilon}^{Z'}_{L_{i}} P_L + {\boldsymbol\epsilon}^{Z'}_{R_{i}} P_R\right){\mathbf F}_i,
 \end{equation}
where $ {\mathbf F}_i$ is a massive fermion triplet in the flavor basis,   $ {\mathbf F}_\ell^T=(e,\mu,\tau)$,  $ {\mathbf F}_d^T=(d,s,b)$, and  $ {\mathbf F}_d^T=(u,c,t)$, with $\boldsymbol{\epsilon}^{Z'}_{L_{i}}$ and $\boldsymbol{\epsilon}^{Z'}_{R_{i}}$ being $3\times 3$ matrices containing the corresponding $Z'$ couplings.
We will focus on the quark up-type sector  since we expect that the largest contribution to TNGBCs arise from the top quark, which will become evident in Sec. \ref{calculations}. As  it was pointed out in Ref. \cite{Arhrib:2006sg}, we assume that the $Z'$ couplings to  down-type quarks $d$, charged leptons $\ell$ and neutrinos $\nu$ are flavor-diagonal and family-universal, namely, $\boldsymbol{\epsilon}^{Z'}_{L_{i},R_{i}}=Q^{i}_{L,R}\mathbf{I}_{3\times 3}$ for $i=d,\ell,\nu$, where $\mathbf{I}_{3\times 3}$ is the  identity matrix and $Q^i_{L,R}$ are the respective chiral charges. As far as the couplings of the $Z'$ gauge boson to  up-type quarks are concerned, we assume that they are family non-universal and are given in the flavor basis as

\begin{equation}
\boldsymbol\epsilon^{Z'}_{L_u}=Q^u_L\left(\begin{array}{ccc}1 & 0 & 0 \\0 & 1 & 0 \\0 & 0 & x\end{array}\right), \quad\quad   \boldsymbol\epsilon^{Z'}_{R_u}=Q^u_R\mathbf{I}_{3\times 3}.
\end{equation}
 Thus, non-universal couplings are only induced through left-handed up-type quarks, with  $x$ a parameter that characterizes  the size of the FCNCs and will be taken as $x\lesssim \mathcal{O}(1)$. The chiral $U'(1)$ charges of the up-type quarks $Q^{u}_{L,R}$ differ in each model as shown in Table \ref{ChiralCharges}.

\begin{table}[hbt!]
\caption{Chiral charges for the models with a new heavy neutral gauge boson of Table \ref{FCNCmodels}. A detailed discussion about the determination of these couplings can be found in Ref. \cite{Arhrib:2006sg}.
\label{ChiralCharges}}
\begin{center}
\begin{tabular}{cccccc}
\hline
\hline
&Sequential $Z$ & $Z_{L,R}$ &$Z_\chi$&$Z_\psi$ & $Z_\eta$ \\
\hline
\hline
$Q^u_L$ & 0.3456 & -0.08493& $\frac{-1}{2\sqrt{10}}$&$\frac{1}{\sqrt{24}}$&$\frac{-2}{2\sqrt{15}}$ \\
$Q^u_R$ & -0.1544 &0.5038& $\frac{1}{2\sqrt{10}}$&$\frac{-1}{\sqrt{24}}$&$\frac{2}{2\sqrt{15}}$ \\
$Q^d_L$ & -0.4228 & -0.08493& $\frac{-1}{2\sqrt{10}}$&$\frac{1}{\sqrt{24}}$&$\frac{-2}{2\sqrt{15}}$ \\
$Q^d_R$ & 0.0772 & -0.6736& $\frac{-3}{2\sqrt{10}}$&$\frac{-1}{\sqrt{24}}$&$\frac{-1}{2\sqrt{15}}$ \\
$Q^e_L$ & -0.2684 & 0.2548&$ \frac{3}{2\sqrt{10}}$&$\frac{1}{\sqrt{24}}$&$\frac{1}{2\sqrt{15}} $\\
$Q^e_R$ & 0.2316 &-0.3339& $\frac{1}{2\sqrt{10}}$&$\frac{-1}{\sqrt{24}}$&$\frac{2}{2\sqrt{15}}$ \\
$Q^\nu _L$ & 0.5 & 0.2548&$ \frac{3}{2\sqrt{10}}$&$\frac{1}{\sqrt{24}}$&$\frac{1}{2\sqrt{15}}$ \\
\hline
\hline
\end{tabular}
\end{center}
\end{table}

After rotating to the mass eigenstates, we obtain the left- and right-handed  up quark fields in the mass eigenbasis via the  $\mathbf{V}_{L_u}$ and $\mathbf{V}_{R_u}$ matrices  respectively. Thus the up-quark term of the Lagrangian of Eq. \eqref{LagrangianZ'} reads
\begin{align}
\label{LagrangianZu}
    \mathcal{L}&=
      -g_{Z'} Z'_\mu \bar{{\mathbf F}}_u^M \gamma_\mu \Big(\mathbf V^\dagger_{L_u} \boldsymbol\epsilon^{Z'}_{L_{u}}\mathbf V_{L_u} P_L \nonumber\\
      &+ \mathbf V^\dagger_{R_u}\boldsymbol\epsilon^{Z'}_{R_{u}}\mathbf V_{R_u} P_R\Big){\mathbf F}_u^M,
   \end{align}
where the superscript $M$ denotes the mass eigenbasis. For simplicity we will drop this superscript below and assume that we are referring to the fermions in the mass eigenstate basis. In general $\mathbf B^u_L\equiv\mathbf V^\dagger_{L_u}\epsilon^{Z'}_{L_u} \mathbf V_{L_u}$ will be non-diagonal. Since no mixing in the down-quark sector is assumed  we will have $\mathbf V_{CKM}=\mathbf V^\dagger_{L_u}\mathbf V_{L_d}=\mathbf V^\dagger_{L_u}$ \cite{Arhrib:2006sg}. Therefore, the flavor mixing will be determined by the CKM matrix:
\begin{align}
\mathbf B^u_L&\equiv\mathbf V_{{CKM}}\boldsymbol\epsilon^{Z'}_{L_u} \mathbf V_{{CKM}}^\dagger\nonumber\\
&\approx \left(\begin{array}{ccc}1 & (x-1)V_{ub}V^\ast_{cb} & (x-1)V_{ub}V^\ast_{tb} \\(x-1)V_{cb}V^\ast_{ub} & 1 & (x-1)V_{cb}V^\ast_{tb} \\ (x-1)V_{tb}V^\ast_{ub} & (x-1)V_{tb}V^\ast_{cb} & x\end{array}\right),
\end{align}
where we have used the unitarity conditions of $\mathbf V_{{CKM}}$. As for  the right-handed couplings   $\mathbf B^u_R\equiv\mathbf V^\dagger_{R_u}\epsilon^{Z'}_{R_u} \mathbf V_{R_u}$, it is easy to see that they  are flavor-diagonal.

The gauge coupling $g_{Z'}$ is the same as that of the SM for the $Z$ gauge boson in the sequential $Z$ model, namely, $g_{Z'}=e/(2s_Wc_W)$, whereas in the  remaining models of Table \ref{FCNCmodels}, it is given by
\begin{equation}
g_{Z'}=\sqrt{\frac{5}{3}}\frac{e}{c_W}\lambda_g^{1/2},
\end{equation}
where $\lambda_g\sim \mathcal{O}(1)$. Below we will assume that $\lambda_g=1$. Constraints on FCNCs arise from $D^0-\overline{D^0}$  mixing \cite{Arhrib:2006sg,Aranda:2010cy}, single top-quark production at the LHC \cite{Gupta:2010wt} and a simple ansatz analysis \cite{He:2009ie}. Implications of  FCNC of a new neutral gauge boson $Z'$  have been studied in leptonic decays of the Higgs boson and the weak bosons \cite{Chiang:2013aha}, $tZ'$ production at the LHC \cite{Hou:2017ozb}, $Z'$ decays \cite{Aranda:2012qs}, $B_s $ and $B_d$ decays \cite{Li:2015xna}, etc.

\section{Analytical results}
\label{calculations}
\begin{figure}[hbt!]
\begin{center}
\subfigure[]{\includegraphics[width=.3\textwidth]{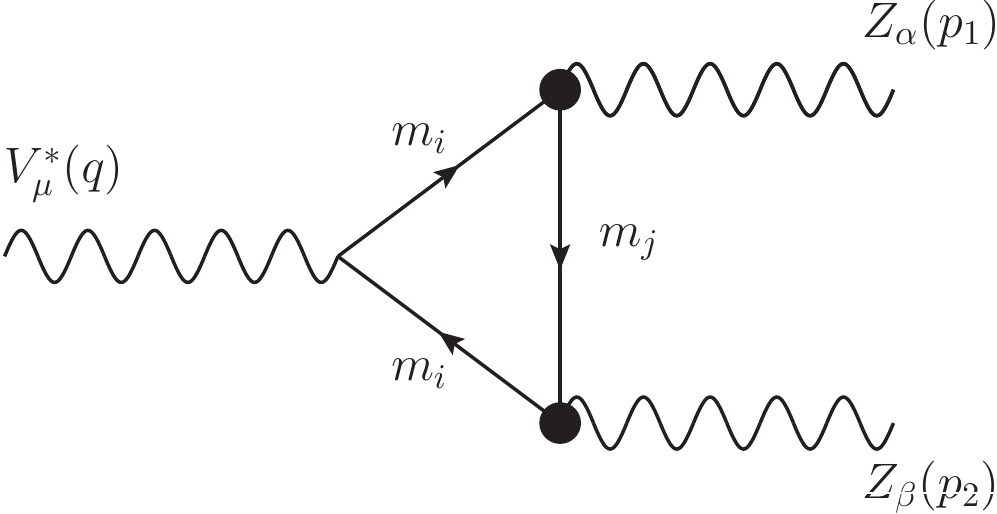}\label{Diagram1}}
\subfigure[]{\includegraphics[width=.3\textwidth]{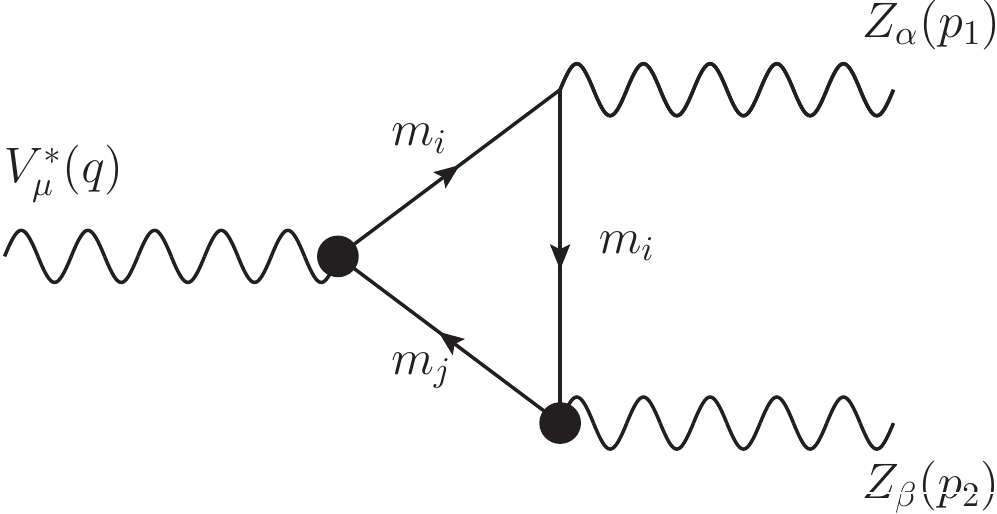}\label{Diagram2}}
\subfigure[]{\includegraphics[width=.3\textwidth]{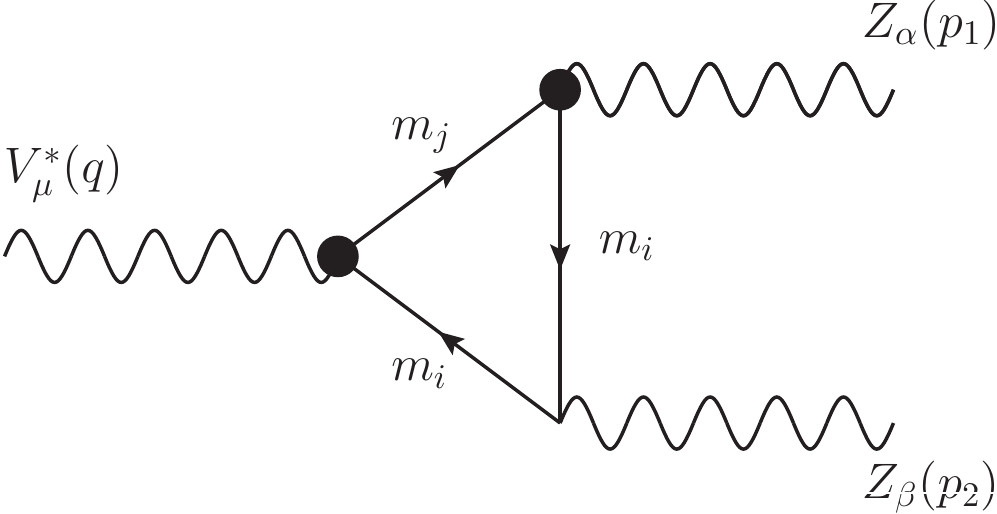}\label{Diagram3}}
\caption{Generic Feynman diagrams required for the contribution of FCNC couplings to TNGBCs $ZZV^\ast$ and $Z\gamma V^\ast$ ($V=Z$,$\gamma$, $Z'$).} \label{FeynmanDiagrams}
\end{center}
\end{figure}

We now turn to present the calculation of the contribution to the TNGBCs  $ZZV^\ast$ ($V=Z$,$\gamma, Z'$) arising from complex  FCNC couplings mediated by the SM $Z$ gauge boson and a new neutral heavy gauge boson $Z'$ as shown in  Eqs. \eqref{LagrangianZ} and \eqref{LagrangianZu}, respectively. This would allow non-vanishing $CP$-violating form factors. For our calculation we will assume conserved vector currents and consider Bose symmetry \cite{Gounaris:2000tb}. This last condition will allows us to obtain all the Feynman diagrams contributing to the TNGBCs.   We will see, however, that  we only need to calculate  the  three generic Feynman diagrams depicted in Fig. \ref{FeynmanDiagrams} since the amplitudes of the additional diagrams  follow easily. For the calculation of the loop amplitudes we use the Passarino-Veltman reduction scheme with the help of the FeynCalc package \cite{Shtabovenko:2016sxi}.

\subsection{$ZZ\gamma^\ast$ coupling}
In this case  there are 4 contributing Feynman diagrams, but  due to gauge invariance we only need to calculate the amplitude of  diagram \ref{Diagram1} ${\cal M}^{\alpha\beta\mu}_2$ since the amplitudes of the remaining diagrams are easily obtained as follows.  There is an additional diagram that is obtained after the exchange $f_i\leftrightarrow f_j$ so its amplitude follows from ${\cal M}^{\alpha\beta\mu}_2$ after  exchanging the  fermion masses ${\cal M}^{\alpha\beta\mu}_2(f_i\leftrightarrow f_j)$. We also need to add a  pair of  diagrams where the $Z$ gauge bosons are exchanged, which means that their total amplitude can be obtained from that of the two already described diagrams after the exchange  $p_{1\mu}\leftrightarrow p_{2\nu}$ is done. We note that it is not possible to induce $CP$ violation in the $ZZ\gamma^\ast$ coupling, which indeed was verified in our explicit calculation. Therefore there are only contributions to the form factor $f_5^\gamma$, which can be written as

\begin{equation}
\label{f5Photon}
f_{5}^\gamma= -\sum_{i}\sum_{j\ne i}\frac{N_i Q_i e^2 m_Z^2  {\rm Re}\left( g_{{AZ}}^{ij\ast} g_{{VZ}}^{ij} \right)}{8 \pi^2 s_W^2 c_W^2 \left( q^2-4m_Z^2\right)^2 q^2} R_{ij},
\end{equation}
where $m_i$, $N_{i}$ and $Q_i$ are the mass, color number and electric charge of the fermion $f_i$. Note that $Q_j=Q_i$ since we are considering neutral currents. The analytical expression for $R_{ij}$  is somewhat cumbersome and  is presented  in  \ref{AnalyticalResults} in terms of Passarino-Veltman scalar functions. We have verified that Eq. \eqref{f5Photon} reduces to that reported  in Ref. \cite{Gounaris:2000tb} for real FCNC couplings of the $Z$ gauge boson.

\subsubsection{Asymptotic behavior}
It is straightforward to obtain the high-energy limit $q^2\gg m_i^2$, $m_j^2$, $m_Z^2$
\begin{equation}
\label{f5photonasym}
f_{5}^{\gamma}\approx-\sum_{i}\sum_{j\ne i} \frac{e^2 Q_i N_i m_Z^2 {\rm Re}\left( g_{{AZ}}^{ij\ast} g_{{VZ}}^{ij} \right)}{4 \pi ^2 q^2
   c_W^2 s_W^2},
\end{equation}
which agrees up to terms of the order $q^{-2}$ with the one  reported in \cite{Gounaris:2000tb} for the $CP$-conserving case ($m_i=m_j$), though we must consider a  factor of $1/2$ as we are counting twice the number of  Feynman diagrams  (our results include the contribution of the diagrams with the exchange $f_i \leftrightarrow f_j$). It is evident that $f_{5}^{\gamma}\rightarrow 0$ in the high-energy limit as  required by unitarity.

In the scenario where an ultra heavy  fermion runs into the loop $m_i^2\gg q^2$, $m_Z^2$, $m_j^2$ must be worked out more carefully as the expansion of the two- and three-point scalar  Passarino-Veltman functions around   small $m_j$ diverge. This scenario could arise in  331 model \cite{Ferreira:2011hm} or little Higgs models \cite{delAguila:2011wk}, for instance, where new heavy quarks and neutrinos are predicted.

\subsection{$ZZZ^\ast$ coupling}
The calculation of this coupling is more intricate than  the previous one since there are 36 contributing  Feynman diagrams, though we
 only need to calculate the three generic Feynman diagrams of Fig. \ref{FeynmanDiagrams}, which by Bose symmetry must be complemented with the diagrams obtained by performing six permutations of four-momenta and Lorentz indices as well as the exchange of the fermions running into the loops.  In this case there are both $CP$-violating and $CP$-conserving form factors. The former is  due to the fact that  the virtual boson is assumed to have complex FCNC couplings. As for the $CP$-conserving form factor $f^{Z}_{5}$, it can be written as

\begin{align}
\label{f5Z}
    f_5^Z&=-\sum_{i}\sum_{j\ne i}\frac{e^2 N_i m_Z^2}{16 \pi ^2 c_W^3 s_W^3 \left(q^2-4 m_Z^2\right)} \nonumber\\
    &\times \Big\{ g^i_{{AZ}} \left(\left| g_{{AZ}}^{ij}\right|
   ^2+\left| g_{{VZ}}^{ij}\right| ^2\right)\left(R_{1ij}+R_{2ij}\right) \\
   &+2g^i_{{VZ}}{\rm Re}\left( g_{{AZ}}^{ij\ast} g_{{VZ}}^{ij} \right)\left(R_{1ij}-R_{2ij}\right)\nonumber\\
   &+g^i_{{AZ}}
   \left[\left| g_{{AZ}}^{ij}\right| ^2-\left|
   g_{{VZ}}^{ij}\right| ^2\right] R_{3ij}+\left( i \leftrightarrow  j\right)  \Big\},
\end{align}
where the $R_{kij}$ ($k=1,2,3$) functions are presented in  \ref{AnalyticalResults} in terms of Passarino-Veltman scalar functions.
This results is in agreement with that reported in Ref. \cite{Gounaris:2000tb} for real FCNC couplings of the $Z$ gauge boson. 

As for the $CP$-violating form factor $f_4^Z$, it reads

\begin{align}
\label{f4Z}
    f_4^Z&=-  \sum_{i}\sum_{j\ne i}\frac{N_i e^2 m_i m_j m_Z^2 }{24\pi
   ^2 c_W^3 s_W^3 \left(q^2-m_Z^2\right) \left(q^2-4  m_Z^2\right)q^2}\nonumber\\
   &\times{\rm Im}\left( g_{{AZ}}^{ij\ast} g_{{VZ}}^{ij} \right)g^i_{{AZ}} S_{ij},
\end{align}
where  $S_{ij}$ is presented in  \ref{AnalyticalResults}. It is easy to see that $f_4^Z$ vanishes for real couplings, which  is also true if we consider  the same fermion running into the loop ($i=j$). Thus, a non-vanishing $f_{4}^Z$ requires complex FCNC couplings. Furthermore, we can also see that we need different complex phases for $g_{{VZ}}^{ij}$ and $g_{{AZ}}^{ij}$ to obtain a non-vanishing $CP$-violating form factor. Since $f_{4}^Z$ is proportional to $m_i m_j$ we expect that the main contribution comes from FCNC couplings associated with the top quark. We would like to stress that the result of Eq. \eqref{f4Z} has never been reported in the literature.

\subsubsection{Asymptotic behavior}
As we did it with the $ZZ\gamma^\ast$ vertex, we study the high-energy limit $q^2\gg m_i^2$, $m_j^2$, $m_Z^2$

\begin{align}
f_{5}^Z&\approx - \sum_{i}\sum_{j\ne i}\frac{e^2 N_i m_Z^2 }{8 \pi ^2 q^2 c_W^3 s_W^3} \Big(g_{{AZ}}^{i}\left( \left| g_{{AZ}}^{ij}\right|^2 +\left| g_{{VZ}}^{ij}\right|^2\right)\nonumber\\
&+2 g_{{VZ}}^{i} {\rm Re}\left( g_{{AZ}}^{ij\ast} g_{{VZ}}^{ij} \right) \Big).
\end{align}
Our result for $f_{5}^Z$ also reproduces the one reported in Ref. \cite{Gounaris:2000tb} for the $CP$-conserving case ($m_i=m_j$), though this time we must consider a factor of 1/6 as we are considering twice the three kinds of Feynman diagrams of Fig. \ref{FeynmanDiagrams} by the exchange $m_i\leftrightarrow m_j$. In ths case we also observe that  $f_{5}\rightarrow 0$ in the high-energy limit, which is consistent with unitarity. The same is true for the $CP$-violating form factor $f_{4}^Z$, which behaves in the high-energy limit as  $ f_4^Z\sim 1/q^4$, since $S_{ij}\sim q^2$ in this limit.

In the case $m_i^2\gg q^2$, $m_Z^2$, $m_j^2$, both form factors diverge similar as in the case of the $ZZ\gamma^\ast$ vertex.

\subsection{$ZZZ^{\prime*}$ coupling}
For the sake of completeness we now consider a  $Z'$ gauge boson with  complex  FCNCs couplings  and calculate the corresponding contributions to the TNGBC $ZZZ^{\prime*}$. We first present the diagonal case, where there is no flavor violation. Since $m_i=m_j$, there is only one independent diagram in Fig. \ref{FeynmanDiagrams} and we only need to add one extra diagram obtained after the exchange  $p_{1\mu}\leftrightarrow p_{2\nu}$. After some algebra, the $CP$-conserving form factor $f_5^{Z'}$  reads
\begin{align}
\label{f5ZpDiagonal}
f_{5}^{Z^\prime}&=-\sum_i\frac{ e N_i m_{Z'}^2}{16 \pi ^2  c_W^2 s_W^2 q^2\left(q^2-4 m_Z^2\right){}^2}\nonumber\\
&\times\Big\{g_{{AZ^\prime}}^i\Big[ (g_{{VZ}}^i)^{2}\ L_{1i} + (g_{{AZ}}^i)^2\ L_{2i} \Big]\nonumber\\
&+g_{{VZ^\prime}}^i\ g_{{VZ}}^i\ g_{{AZ}}^i\ L_{3i}\Big\},
\end{align}
where  the $L_{ji}$ ($j=1$, 2, 3) functions are presented in  \ref{AnalyticalResults}. The $CP$-violating form factor $f_4^{Z'}$ is not induced at the one-loop level in this scenario.

As far as the non-diagonal case with complex FCNCs couplings, it requires more effort. Apart from the three generic Feynman diagram of Fig. \ref{FeynmanDiagrams}, we must add those diagrams obtained after  the exchanges $p_{1\mu}\leftrightarrow p_{2\nu}$ and $f_1 \leftrightarrow f_2$, so  there are 12 contributing Feynman diagrams in total. However,  we only need to calculate the amplitudes of the three generic diagrams. In this scenario both $f_{5}^{Z'}$ and $f_{4}^{Z'}$ form factors are non-vanishing. The $CP$-conserving form factor can be written as

\begin{align}
\label{f5Zp}
f_5^{Z^\prime}&= - \sum_{i}\sum_{j\ne i}\frac{e N_i m_{Z'}^2}{16 \pi ^2  c_W^2 s_W^2  q^2 \left(q^2-4 m_Z^2\right)^2} \nonumber\\
&\times \Big\{ 2g_{{AZ}}^i\Big[ {\rm Re} \left( g_{VZ}^{ij} g_{VZ'}^{ij\ast} \right) U_{1ij} \nonumber\\
&+ {\rm Re} \left( g_{AZ'}^{ij} g_{AZ}^{ij\ast} \right) U_{2ij}\Big]+2g_{VZ'}^i {\rm Re} \left( g_{VZ}^{ij} g_{AZ}^{ij\ast} \right)U_{3ij}  \nonumber\\
&  +2 g_{{VZ}}^i\Big[ {\rm Re} \left( g_{VZ}^{ij} g_{AZ'}^{ij\ast} \right) U_{4ij}+{\rm Re} \left( g_{VZ'}^{ij} g_{AZ}^{ij\ast} \right) U_{5ij} \Big]\nonumber \\
   &+g_{AZ'}^i\Big[\left| g_{{AZ}}^{ij}\right|
   ^2  U_{6ij}+\left| g_{{VZ}}^{ij}\right| ^2 U_{7ij}\Big]\Big\},
\end{align}
whereas the $CP$-violating one reads
\begin{align}
\label{f4Zp}
f_4^{Z^\prime}&=  \sum_{i}\sum_{j\ne i}\frac{ e N_i m_{Z^\prime}^2}{12 \pi ^2 c_W^2 s_W^2 q^6  \left(q^2-4
   m_Z^2\right)} \nonumber\\
   &\times \Big\{ g_{VZ}^{i}\Big[ {\rm Im} \left( g_{VZ'}^{ij} g_{VZ}^{ij\ast} \right) T_{1ij}+ {\rm Im} \left( g_{AZ'}^{ij} g_{AZ}^{ij\ast} \right) T_{2ij}\Big]\nonumber\\
   &+ g_{{AZ}'}^i{\rm Im} \left( g_{VZ}^{ij} g_{AZ}^{ij\ast} \right) T_{3ij}  \nonumber\\
   &+ g_{AZ}^i\Big[ {\rm Im} \left( g_{AZ'}^{ij} g_{VZ}^{ij\ast} \right) T_{4ij} +{\rm Im} \left( g_{VZ'}^{ij} g_{AZ}^{ij\ast} \right) T_{5ij}\Big] \Big\},
\end{align}
where the $U_{kij}$ ($k=1\ldots 7$) and $T_{kij}$ ($k=1\ldots 5$) functions are presented in  \ref{AnalyticalResults}. We note that
$f_5^{Z^\prime}$ ($f_4^{Z^\prime}$) depends only  on the  real (imaginary) part of the  combinations of products of the vector and axial couplings. We also note that it is not necessary that both $Z$ and $Z'$ gauge bosons have simultaneously complex FCNC couplings to induce the $CP$-violating form factor.

\subsubsection{Asymptotic behavior}

In the diagonal case  the form factor $f_5^{Z'}$ can be written in the high-energy limit $q^2\gg m_i^2$, $m_j^2$, $m_Z^2$ as
\begin{align}
\label{f5asy1}
f_5^{	Z'}&\simeq-\sum_i\frac{e^2 m_{Z'}^2 N_i }{32 \pi ^2  c_W^3 s_W^3 q^2}\Big\{g_{{AZ'}}^i \left((g_{{AZ}}^i)^2+(g_{{VZ}}^i)^2\right)\nonumber\\
&+2
   g_{{AZ}}^i\ g_{{VZ}}^i\ g_{{VZ^\prime}}^i\Big\},
\end{align}
whereas in the non-diagonal case we obtain
\begin{align}
\label{f5asy2}
   f_5^{Z'} &\simeq -\sum_{i}\sum_{j\ne i}\frac{e^2 m_{Z'}^2 N_i }{16 \pi ^2  c_W^3 s_W^3 q^2}\nonumber\\
&\times\Big\{g_{AZ'}^i \left(\left| g_{{AZ}}^{ij}\right|
   ^2+\left| g_{{VZ}}^{ij}\right|
   ^2\right)\nonumber\\
&+2  \Big[ g_{AZ}^i\Big( {\rm Re}\left(g_{AZ'}^{ij}g_{AZ}^{ij\ast} \right)+{\rm Re}\left(g_{VZ}^{ij}g_{VZ'}^{ij\ast} \right)\Big)\nonumber\\
   &+g_{VZ}^i \Big({\rm Re} \left( g_{VZ'}^{ij} g_{AZ}^{ij\ast} \right)+{\rm Re} \left( g_{VZ}^{ij} g_{AZ'}^{ij\ast} \right)\Big)\nonumber\\
&+g_{VZ'}^i {\rm Re} \left( g_{VZ}^{ij} g_{AZ}^{ij\ast} \right)\Big]
   \Big\} .
\end{align}
We note that in both scenarios $f_5^{Z'}\sim m_{Z'}^2/q^2$, thereby decreasing quickly when $q^2\geqslant m_{Z'}^2$. However this effect is attenuated for $q^2\lesssim m_{Z'}^2$ due to  the mass of the heavy $Z'$ boson.  We also note that  Eq. \eqref{f5asy2} reduces to Eq. \eqref{f5asy1} except by a factor of 6, which is due to the fact that the $CP$-violating form factor receives the contribution of twelve Feynman diagrams in the diagonal scenario instead of  two as in the diagonal case.

On the other hand, the $CP$-violating form factor $f_4^{Z'}$ is of the order of  $m_{Z'}^2/q^4$ in the high energy limit and decreases quickly as $q^2$ increases. The functions $T_{kij}$ behave in this limit as $T_{kij}\sim q^4$, therefore the form factor $f_4^{Z'}$ as functions of $q^2$ has the form  $f_4^{Z'}\sim 1/q^4$, which is similar to the vertex $ZZZ^\ast$ in the high energy limit.

Furthermore, when  $m_i^2\gg q^2$, $m_Z^2$, $m_j^2$ both form factors also show the same behavior  observed in the case of the vertices $ZZ\gamma^\ast$ and $ZZZ^\ast$.

\section{Constraints on FCNC $Z$ couplings}
\label{SectionConstrains}
 We would like to assess  the magnitude of the new contributions to the $f_4^{V}$ and $f_5^{V}$ ($V=\gamma, Z, Z'$) form factors. It is thus  necessary to obtain constraints on the FCNC $Z$ couplings to obtain an estimate of the numerical values of such form factors. Since we expect that the main contributions arise from the FCNC couplings of the top quark, we use the current bounds on the branching ratios  of the FCNC decays $t\rightarrow qZ$, where $q=c$, $u$ \cite{Aaboud:2018nyl} to constrain the $g_{{VZ},{AZ}}^{tq}$ couplings.

\subsection{Constraints on the FCNC $Z$ couplings from $t\rightarrow q Z$ decay}
A comprehensive compilation of the branching ratios of top FCNC decays within the SM and several extension models can be found in \cite{AguilarSaavedra:2004wm}. In the case of tree-level  FCNC $Z$ couplings,  the decay width $t\rightarrow qZ$ can be written  in terms of the vector and axial couplings for negligible $m_q$  as follows
\begin{align}
\Gamma_{t\rightarrow Zq}&=\frac{e^2 m_t^3}{64 \pi  c_W^2  m_Z^2 s_W^2}\left(\left| g_{{AZ}}^{tq}\right| ^2+\left|g_{{VZ}}^{tq}\right| ^2\right) \nonumber\\
&\times\left(1-\frac{m_Z^2}{m_t^2}\right)^2
   \left(1+2 \frac{m_Z^2}{m_t^2}\right).
\end{align}
The current upper limits obtained by ATLAS collaboration at $\sqrt{s}=$ 13 TeV are: $\mathcal{B}(t\rightarrow uZ)< 1.7\times 10^{-4}$ and $\mathcal{B}(t\rightarrow cZ)< 2.4\times 10^{-4}$  with 95\% C.L. \cite{Aaboud:2018nyl}. Previous results at $\sqrt{s}=7$ TeV are also available \cite{Aad:2012ij}.  The SM contribution to the $t\rightarrow cZ$ branching ratio is negligible: $\mathcal{B}\left( t\rightarrow cZ\right)\simeq  10^{-14}$ \cite{AguilarSaavedra:2004wm}. We thus obtain for the contribution of the FCNC couplings of the $Z$ gauge boson:
\begin{equation}
\mathcal{B}\left(t\rightarrow qZ \right)=0.915699 \left(\left| g_{{AZ}}^{tq}\right| ^2+\left|g_{{VZ}}^{tq}\right| ^2\right),
\end{equation}
which allows us to obtain the following limits
\begin{equation}
\label{tuBound}
\left| g_{{AZ}}^{tu}\right| ^2+\left|g_{{VZ}}^{tu}\right| ^2< 1.8\times10^{-4},
\end{equation}
and
\begin{equation}
\label{tcBound}
\left| g_{{AZ}}^{tc}\right| ^2+\left|g_{{VZ}}^{tc}\right| ^2<2.6\times 10^{-4}.
\end{equation}
Eqs. \eqref{tuBound} and \eqref{tcBound} can also be written in terms of the chiral couplings of Eq. \eqref{chiral}.

\begin{figure*}[!htb]
\begin{center}
\includegraphics[width=.8\textwidth]{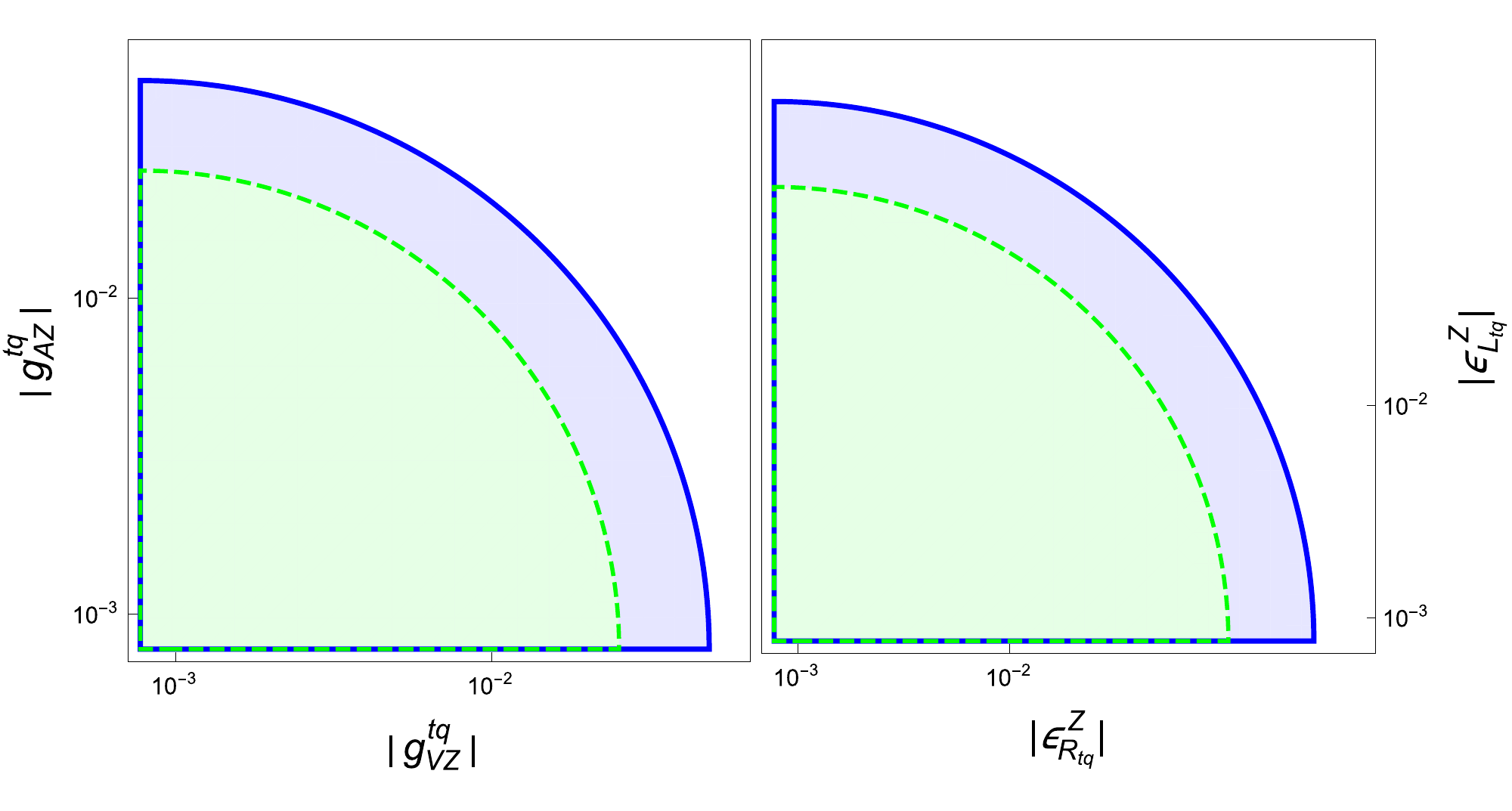}
\caption{Allowed area with 95\% C.L. in the $\left| g_{{AZ}}^{tq}\right|$ vs $\left|g_{{VZ}}^{tq}\right| $ (left) and $|\epsilon _{R_{tq}}^Z|$ vs $|\epsilon _{L_{tq}}^Z|$ (right) planes from the experimental bounds on $t\rightarrow Zq$ decays, for the  $Z\overline{t}c$ and (solid-line boundaries) and $Z\overline{t}u$ (dashed-line boundaries) couplings. }\label{plotBounds}
\end{center}
\end{figure*}

We show in Fig. \ref{plotBounds} the allowed areas on the  $\left| g_{{AZ}}^{tq}\right|$ vs $\left|g_{{VZ}}^{tq}\right| $ and $|\epsilon _{R_{tq}}^Z|$ vs $|\epsilon _{L_{tq}}^Z|$ planes. The blue-solid (green-dashed) line  corresponds to the $Z\overline{t}c$  ($Z\overline{t}u$) couplings. We observe that the FCNC couplings of the $Z$ gauge boson can be as large as $10^{-1}$.  In fact, if we assume $\left| g_{{AZ}}^{tq}\right|  \simeq \left|g_{{VZ}}^{tq}\right| $, we obtain
\begin{equation}
\left|g_{{VZ}}^{tu}\right|< 0.0096,\quad\quad \left|g_{{VZ}}^{tc}\right|<0.011,
\end{equation}
which in terms of the chiral coupling read
\begin{equation}
|\epsilon _{R_{tu}}^Z|<0.013, \quad\quad |\epsilon _{R_{tc}}^Z|<0.016.
\end{equation}
Thus, our bounds are of  the order of  $10^{-2}-10^{-3}$, which are similar to the constraints on FCNC couplings of down quarks obtained from $B$ and Kaon meson decays. For instance the constraint on the $\left|g_{{VZ}}^{bd}\right|$ coupling is at the $ 10^{-2}-10^{-3}$ level \cite{Buchalla:2000sk,Mohanta:2005gm,Giri:2003jj,Silverman:1991fi,Buras:1998ed}, whereas the $\left|g_{{VZ}}^{bs}\right|$ coupling is constrained to be below  $10^{-1}$ \cite{Buchalla:2000sk}. In some extension  models  these couplings can be of the order of $10^{-4}-10^{-7}$ \cite{Colangelo:1998pm,Nir:1990yq}.

\subsection{Constraints on the lepton flavor violating $Z$ couplings from $Z\rightarrow \ell_i\ell_j$}
Following the above approach, we now obtain constraints on the lepton flavor violating (LFV) couplings of the $Z$ gauge boson from the experimental limits  on the $Z\rightarrow \ell^{\pm}\ell^{\mp}$ decays, which have been obtained by  the ATLAS and CMS collaborations: $\mathcal{B}\left(Z\rightarrow e\tau \right)<5.8 \times10^{-5}$, $\mathcal{B}\left(Z\rightarrow \mu\tau \right)<2.4\times 10^{-5}$  at $\sqrt{s}=14$ TeV \cite{Aaboud:2018cxn} and $\mathcal{B}\left(Z\rightarrow e\mu \right)< 7.3\times 10^{-7}-7.5\times10^{-7}$ at $\sqrt{s}=8$ TeV  \cite{Nehrkorn:2017fyt,Aad:2014bca}. The decay width $Z\rightarrow \ell_i\ell_j$ is  given by
\begin{equation}
\Gamma_{Z\rightarrow \ell_i\ell_j}=\frac{e^2 m_Z  }{24 \pi  c_W^2
   s_W^2} \left(\left| g_{{AZ}}^{\ell_i \ell_j}\right| ^2+\left|g_{{VZ}}^{\ell_i \ell_j}\right| ^2\right),
\end{equation}
and the corresponding  branching ratio is
\begin{equation}
\mathcal{B} \left(Z\rightarrow \ell_i\ell_j \right)= 0.250277 \left(\left| g_{{AZ}}^{\ell_i \ell_j}\right| ^2+\left|g_{{VZ}}^{\ell_i \ell_j}\right| ^2\right).
\end{equation}
If  we assume  that $\left| g_{{AZ}}^{\ell_i \ell_j}\right| \simeq \left|g_{{VZ}}^{\ell_i \ell_j}\right| $, we obtain with 95 \% C.L.
 \begin{equation}
\left|g_{{VZ}}^{\tau \mu}\right|<0.0069, \quad{}\quad \left|g_{{VZ}}^{\tau e}\right|< 0.01, \quad{}\quad    \left|g_{{VZ}}^{\mu e}\right|<0.0012.
\end{equation}
We note that the constraints on $\left|g_{{VZ}}^{\tau e}\right|$ and $\left|g_{{VZ}}^{\mu e}\right|$ are less competitive to those obtained through the $\mu\rightarrow eee$ and $\tau^{-}\rightarrow e^-\mu^+\mu^-$ decays \cite{Mohanta:2010yj}, which yield $\left|g_{{VZ}}^{\tau e}\right|< 1.28\times 10^{-3}$ and $\left|g_{{VZ}}^{\mu e}\right|<3.05\times 10^{-6}$. As for the bound on $\left|g_{{VZ}}^{\tau \mu}\right|$, it  is of the same order than the one obtained from the $\tau^-\rightarrow \mu^-\mu^+\mu^-$ decay, namely, $\left|g_{{VZ}}^{\tau \mu}\right|<1.295 \times 10^{-3}$ \cite{Mohanta:2010yj}. In our analysis below we consider the most stringent bounds, thus we will use the values reported in Ref. \cite{Mohanta:2010yj}.

As far as the LFV $Z$ couplings to neutrinos are concerned,  there are no  experimental data to obtain reliable constraints, so to obtain a rough estimate of these contributions we can assume couplings of the same order of magnitude than those used for the charged leptons. Nevertheless, the $f_4^{V}$ and $f_5^{V}$ form factors  are mainly dominated by the contribution of the heaviest quarks, whereas the lepton contributions are negligibly.

Finally, in Table \ref{TableConstrainsZf1f2} we summarize the constraints on the FCNC $Z$ gauge boson couplings that we will use in our numerical analysis, in terms of the corresponding chiral couplings.

 \begin{table*}[hbt!]
\caption{Bounds on the FCNC couplings  of the $Z$ gauge boson, with 95 \% C.L., from the current experimental limits on FCNC $Z$ decays. The second row stands for  the limit when either
$\epsilon_{R_{ij}}^Z$ or $ \epsilon_{L_{ij}}^Z$ is taken as vanishing and the other one non-vanishing.
}
\label{TableConstrainsZf1f2}
\begin{center}
\begin{tabular}{ccccccc}
\hline
\hline
  &$\overline{t}c$&$\overline{t}u$&$\overline{c}u$&$\overline{d}_id_j$&$\overline{\ell}_i\ell_j$&$\overline{\nu}_i\nu_j$\\
\hline
\hline
 $\left|\epsilon_{L_{ij}}^Z\right|\simeq \left|\epsilon_{R_{ij}}^Z\right|$ &0.016&0.013&$10^{-2}$&$10^{-2}$&$10^{-3}$&$10^{-3}$ \\
 $\left|\epsilon_{L_{ij}, R_{ij}}^Z\right|$   &0.032&0.026&$10^{-2}$&$10^{-2}$&$10^{-3}$&$10^{-3}$ \\
\hline
\hline
\end{tabular}
\end{center}
\end{table*}

\section{Numerical Analysis }
We now turn to present the numerical evaluation of the TNGBCs. For the numerical analysis we evaluate the Passarino-Veltman scalar functions via the LoopTools \cite{Hahn:1998yk} package and independently by the Collier \cite{Denner:2016kdg} package, which give a good agreement. We first analyze the case of the $ZZV^*$ ($V=\gamma,Z$) couplings. As a matter of convenience, we write the  complex chiral FCNC $Z$ couplings as
\begin{equation}
\epsilon^{Z}_{L_{ij},R_{ij}}=\overline\epsilon^{Z}_{L_{ij},R_{ij}} +i \tilde\epsilon^{Z}_{L_{ij},R_{ij}},
\end{equation}
where the bar (tilde) denotes the real (imaginary) part of each coupling. The $CP$-violating phase can then be written as

\begin{equation}
\arctan\left(\phi_{L_{ij},R_{ij}}\right)=\frac{\tilde\epsilon^{Z}_{L_{ij},R_{ij}}}{\overline\epsilon^{Z}_{L_{ij},R_{ij}}}.
\end{equation}
We thus can write the real and imaginary terms that enter into the $f_4^{V}$ and $f_5^{V}$ ($V=\gamma, Z, Z'$) form factors [Eqs. \eqref{f5Photon}-\eqref{f4Z}] as follows

\begin{equation}
\left| g_{{VZ},AZ}^{ij}\right| ^2=\frac{1}{4}\left( \left(  \overline\epsilon^{Z}_{L_{ij}}\pm\overline\epsilon^{Z}_{R_{ij}}\right)^2
+\left( \tilde\epsilon^{Z}_{L_{ij}}\pm\tilde\epsilon^{Z}_{R_{ij}}\right)^2 \right),
\end{equation}

\begin{align}
2{\rm Re}\left( g_{{AZ}}^{ij\ast} g_{{VZ}}^{ij} \right)&=\frac{1}{2}\Big(\left(\overline\epsilon^{Z}_{L_{ij}}\right)^2 - \left(\overline\epsilon^{Z}_{R_{ij}}\right)^2+\left(\tilde\epsilon^{Z}_{L_{ij}}\right)^2\nonumber\\
&- \left(\tilde\epsilon^{Z}_{R_{ij}}\right)^2\Big),
\end{align}

\begin{equation}
2{\rm Im}\left( g_{{AZ}}^{ij\ast} g_{{VZ}}^{ij} \right)=\left(\overline\epsilon^{Z}_{L_{ij}} \tilde\epsilon^{Z}_{R_{ij}}-\overline\epsilon^{Z}_{R_{ij}}\tilde\epsilon^{Z}_{L_{ij}}   \right).
\end{equation}
Below we will  analyze the behavior of the $f_{4,5}^{V}$ form factors as functions of the $\overline\epsilon^{Z}_{L_{ij},R_{ij}}$ and $\tilde\epsilon^{Z}_{L_{ij},R_{ij}}$ parameters as well as the transfer momentum $q^2$ of the $V$ gauge boson.

\subsection{$ZZ\gamma^\ast$ coupling}

It is convenient to  assume small phases  of the FCNC $Z$ couplings, namely, we consider that the imaginary parts of the left-handed couplings are smaller than ten percent of their real parts:
\begin{equation}
\phi_{L_{ij}}\simeq\frac{\tilde\epsilon^{Z}_{L_{ij}}}{\overline\epsilon^{Z}_{L_{ij}}}\le  O\left(10^{-1}\right),
\end{equation}
whereas for the right-handed couplings we assume by simplicity that $\phi_{R_{ij}}=0$ ($\tilde\epsilon^{Z}_{R_{ij}}=0$).
As far as the size of the  chiral couplings $\tilde\epsilon^{Z}_{L_{ij},R_{ij}}$ we consider the bounds shown in Table \ref{TableConstrainsZf1f2} to obtain an estimate of  $f_5^\gamma$, which is the  only   non-vanishing $ZZ\gamma^*$ form factor.

\begin{figure*}[!hbt]
\begin{center}
\includegraphics[width=10cm]{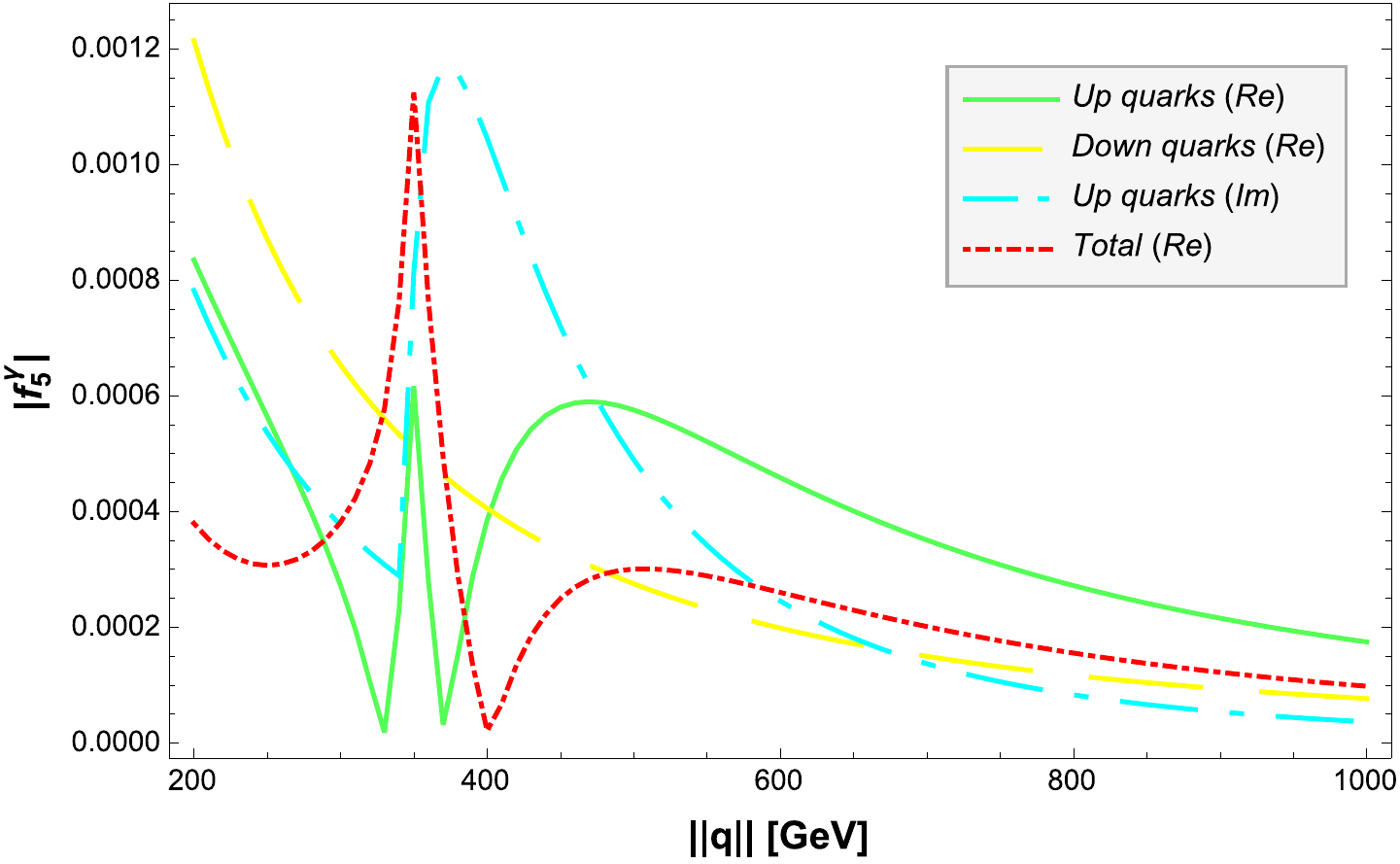}
\caption{Behavior of the FCNC contributions to the $f_5^\gamma$ form factor as a function of the momentum of the photon for $\phi_{L_{ij}}=0.1$, $\phi_{R_{ij}}=0$,
$|\epsilon^{Z}_{R_{ij}}|=0.9|\epsilon^{Z}_{L_{ij}}|$ and the $|\epsilon^{Z}_{L_{ij}}|$ values shown in Table \ref{TableConstrainsZf1f2}.   Only the non-negligible contributions are shown: up quarks ($tc$, $tu$ and $cu$) and down quarks ($bs$, $bd$ and $sd$). The total imaginary contribution coincides with the respective up quark contribution since the down quark contribution (not shown in the plot) is negligible.\label{plotf5photon}}%
\end{center}
\end{figure*}

We show the behavior of  the FCNC contributions to $f_5^\gamma$ as a function of the photon transfer momentum $q^2$  in Fig. \ref{plotf5photon}, where we only plot the non-negligible imaginary and real parts arising from each fermion loop as well as their total sum. We find that the only non-negligible  contributions arise from the up and down quarks, though the former are the only ones yielding a non-negligible imaginary part, which thus coincides with the total imaginary contribution.   We  have considered the parameter values of Table \ref{TableConstrainsZf1f2}, but the curves shown in Fig. \ref{plotf5photon} exhibit a similar behavior for other parameter values: there is a shift upwards (downwards) when the chiral couplings values increase (decrease) as $f_5^\gamma$ is proportional to ${\rm Re}\left( g_{{AZ}}^{ij\ast} g_{{VZ}}^{ij} \right)$. We then conclude that the contributions to $f_5^\gamma$ arising from FCNCs $Z$ couplings are expected to be considerably smaller than the SM contribution, which is of the order of $10^{-2}$.

\subsection{$ZZZ^\ast$ coupling}

\begin{figure*}[!hbt]
\begin{center}
\includegraphics[width=18cm]{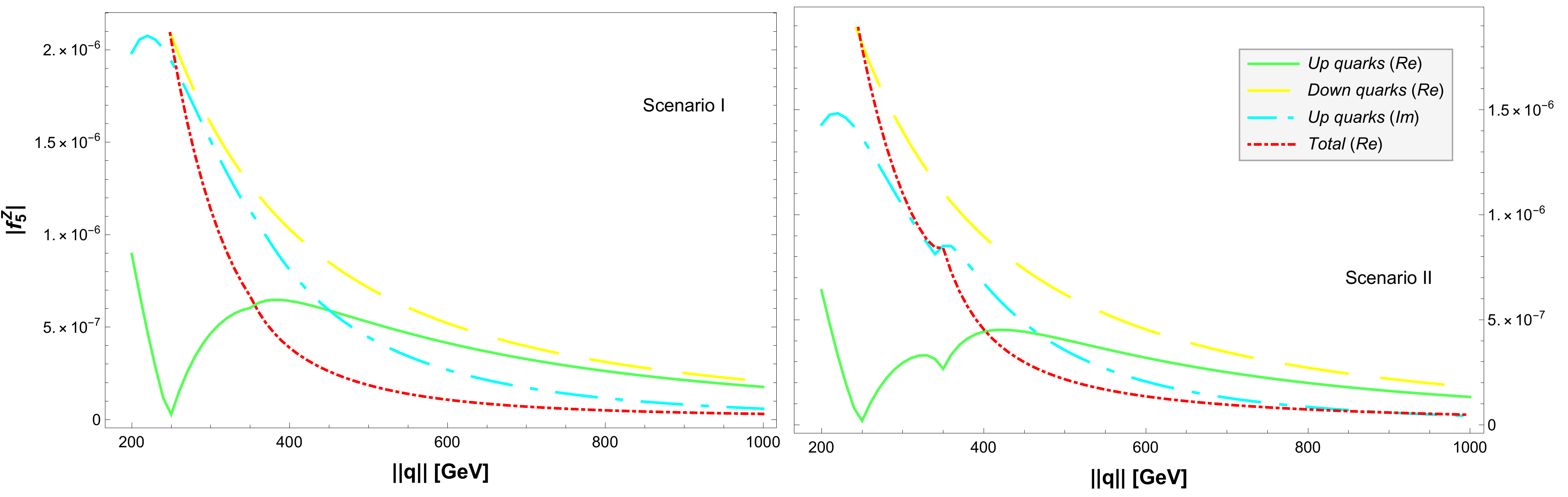}
\caption{Behavior of the FCNC contributions to the $f_5^Z$ form factor as a function of the momentum of the virtual $Z$ gauge boson in the two scenarios discussed in the text.   Only the non-negligible contributions are shown: up quarks ($tc$, $tu$ and $cu$) and down quarks ($bs$, $bd$ and $sd$). The total imaginary contributions coincide with the respective up quark contributions since the down quark contribution (not shown in the plot) is negligible.
\label{plotf5Z}}%
\end{center}
\end{figure*}

In this case both $f_4^Z$ and $f_5^Z$ are non-vanishing.  For our analysis we find it convenient to consider two scenarios:

\begin{itemize}
  \item Scenario I (Left- and right-handed couplings of similar size):  $|\overline\epsilon^{Z}_{R_{ij}}|=0.9|\overline\epsilon^{Z}_{L_{ij}}|$, $\phi_{L_{ij}}=0.1$, and $\phi_{R_{ij}}=0$.
  \item Scenario II (Dominating left-handed couplings): \\ $|\overline\epsilon^{Z}_{R_{ij}}| \simeq 10^{-1}\times |\overline\epsilon^{Z}_{L_{ij}}|$, $\phi_{L_{ij}}=0.1$, and $\phi_{R_{ij}}=0$.
\end{itemize}
We do not consider the  scenario with dominating right-handed couplings since there is no substantial change in the magnitude of the $ZZZ^*$ form factors as that observed in Scenario II. We show in Fig. \ref{plotf5Z} the behavior of the FCNC contributions to $f_5^Z$ as a function of the virtual $Z$ transfer momentum $q^2$ in the two scenarios described above. Again we only show the real and imaginary parts arising from the  up and down quarks along with the total contribution, though  the imaginary part of the down quark contribution is negligible and is not shown in the plots.  We observe that the largest values of $f_5^Z$ are of the order of $10^{-6}$, which are reached for smaller $q^2$ but decrease by one order of magnitude as $q^2$ becomes large. Again, the contribution to $f_5^Z$ from FCNC $Z$ couplings is smaller than the SM contribution.

\begin{figure*}[!hbt]
\begin{center}
\includegraphics[width=18cm]{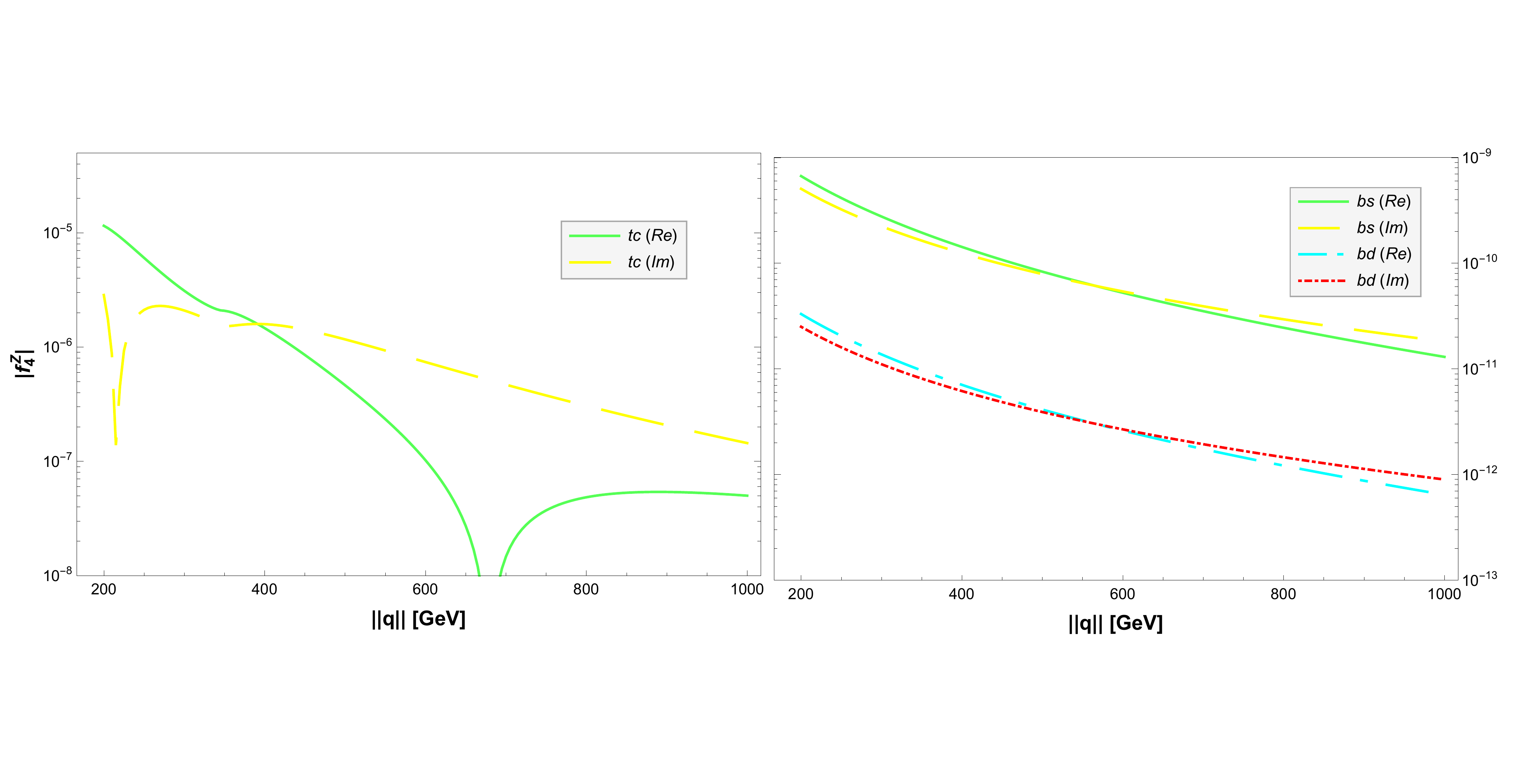}
\caption{Behavior of the $f_4^Z$ form factor as a function of the momentum $|q|$ of the virtual $Z$ gauge boson.  We have extracted a factor of ${\rm Im}\left( g_{{AZ}}^{tc\ast} g_{{VZ}}^{qq'} \right)$ from the respective contribution. All other contributions not shown in the plots are well below the $10^{-10}$ level. \label{plotf4Z}}%
\end{center}
\end{figure*}

We now analyze the $f_4^Z$ form factor, which is absent in the SM up to the one-loop level, which means that any sizeable excess can be attributed to new physics effects. We find that the only non-negligible contributions to $f_4^Z$ arise from the loops induced by the $Ztc$ coupling, with the remaining up and down quark contributions being several orders of magnitude smaller. We thus show in the left plot of Fig. \ref{plotf4Z} the $Ztc$ contribution to the $f_4^Z$ form factor as a function of the virtual $Z$ four-momentum, whereas in the right plot we show the $Zbs$ and $Zbd$  contributions. We have extracted the factor  ${\rm Im}\left( g_{{AZ}}^{qq'\ast} g_{{VZ}}^{qq'} \right)$, so for the $Ztc$ contribution $f_4^Z$ is of the order of
\begin{equation}
|f_4^Z|\simeq |{\rm Im}\left( g_{{AZ}}^{tc\ast} g_{{VZ}}^{tc} \right)|\times 10^{-5},
\end{equation}
for relatively small $||q||\sim 200 $ GeV, but there is a decrease of up to two orders of magnitude as $||q||$ becomes of the order of a few TeVs. All the remaining contributions are considerably suppressed.

\subsection{$ZZZ^{'\ast}$ coupling}

We now turn to the analysis of  the $CP$-conserving $f_{5}^{Z'}$ and   the $CP$-violating $f_4^{Z'}$ form factors for an off-shell $Z'$ boson. For the FCNCs couplings mediated by the $Z$ gauge boson we consider the same scenarios analyzed in the case of the $ZZZ^\ast$ vertex. We thus use the constraints presented  in Table \ref{TableConstrainsZf1f2}, which were obtained from the data on $Z\overline{f}_if_j$ decays. Furthermore, for the $Z'$ couplings we use the interaction of Eq. \eqref{LagrangianZ'}, with the values of Table \ref{ChiralCharges} for the chiral charges, along with  $x=0.1$ and ${\phi_{L^\prime}}_{ij}=0.001$. Here  $x$ stands for   the parameter characterizing the size of $Z'$ FCNC couplings and ${\phi_{L^\prime}}_{ij}$ is the $CP$-violating phase of the $Z'$ couplings to left-handed up quarks. Since all the models summarized in Table \ref{ChiralCharges} give rise to
 similar results, we will only present the numerical results for the $Z_\eta$ model.

\begin{figure*}[!hbt]
\begin{center}
\includegraphics[width=18cm]{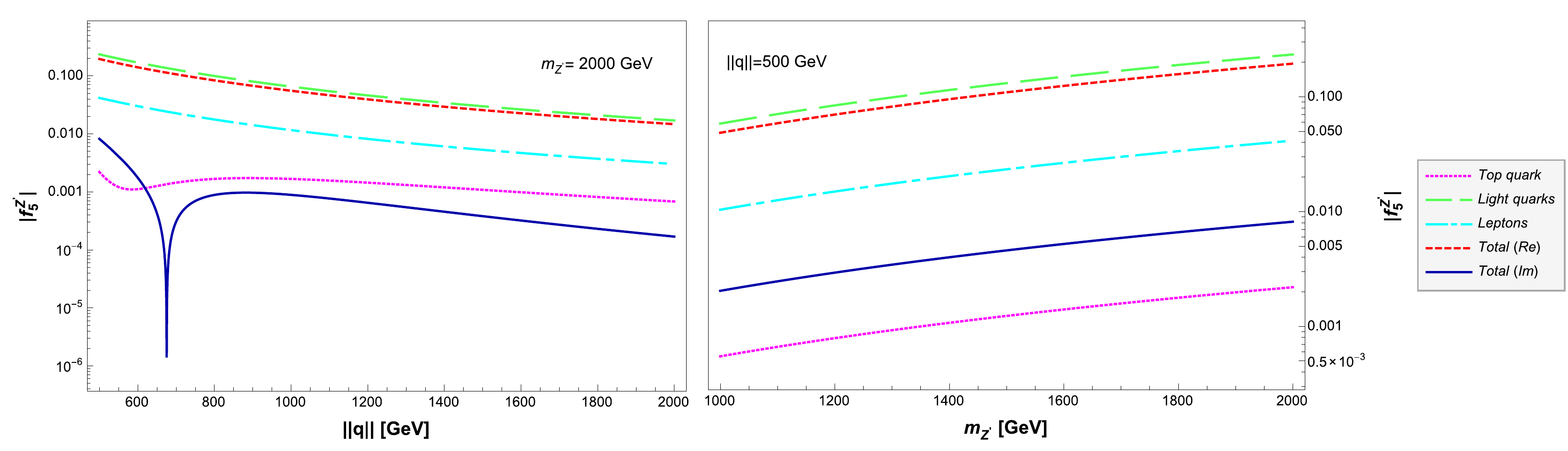}
\caption{Behavior of the $f_5^{Z^\prime}$ form factor as a function of the transfer momentum $|q|$ (left plot) and the $m_{Z^\prime}$ mass (right plot) in the diagonal case .   \label{plotZp1}}%
\end{center}
\end{figure*}

\begin{figure*}[!hbt]
\begin{center}
\includegraphics[width=7cm]{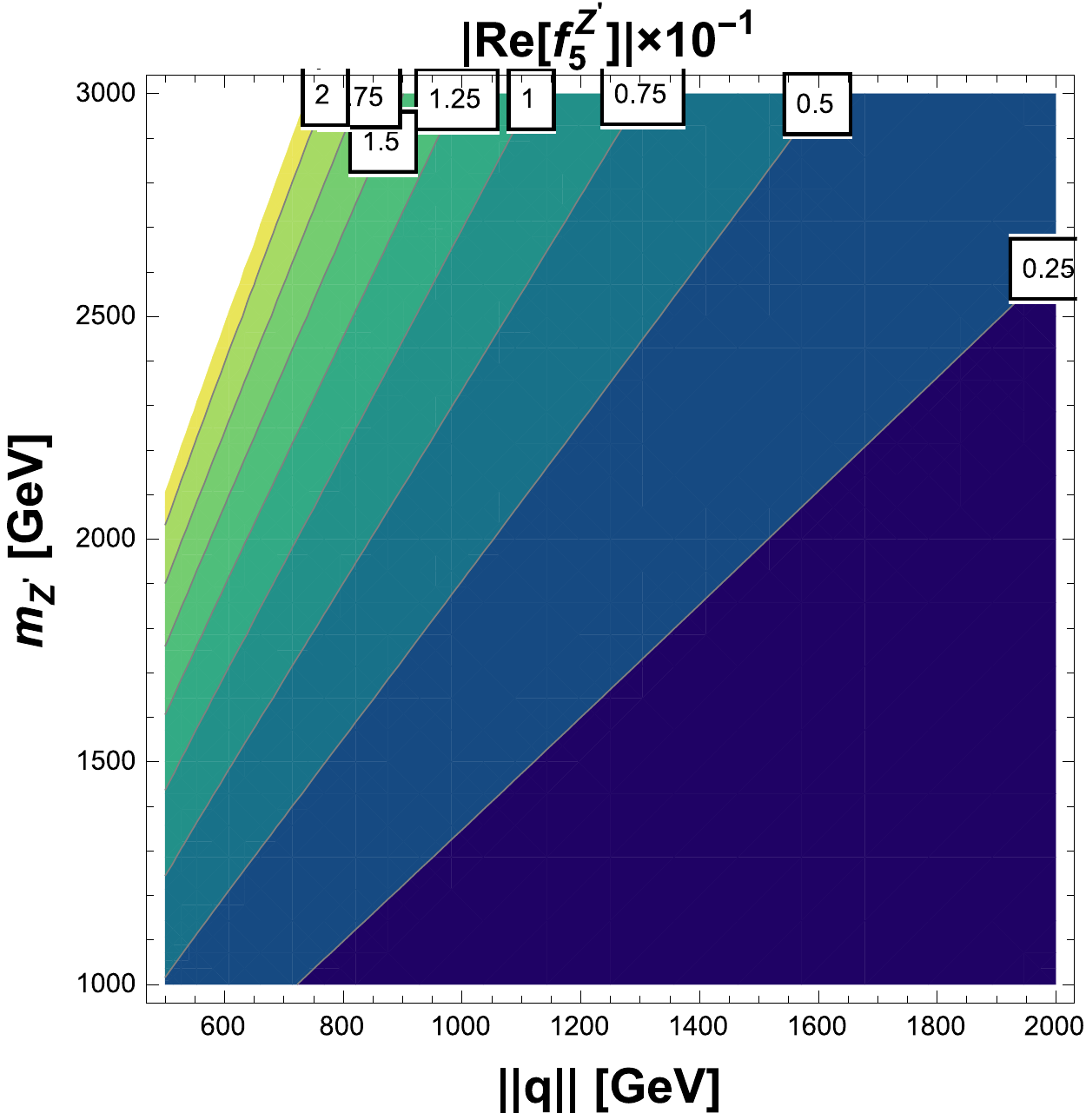}
\includegraphics[width=7cm]{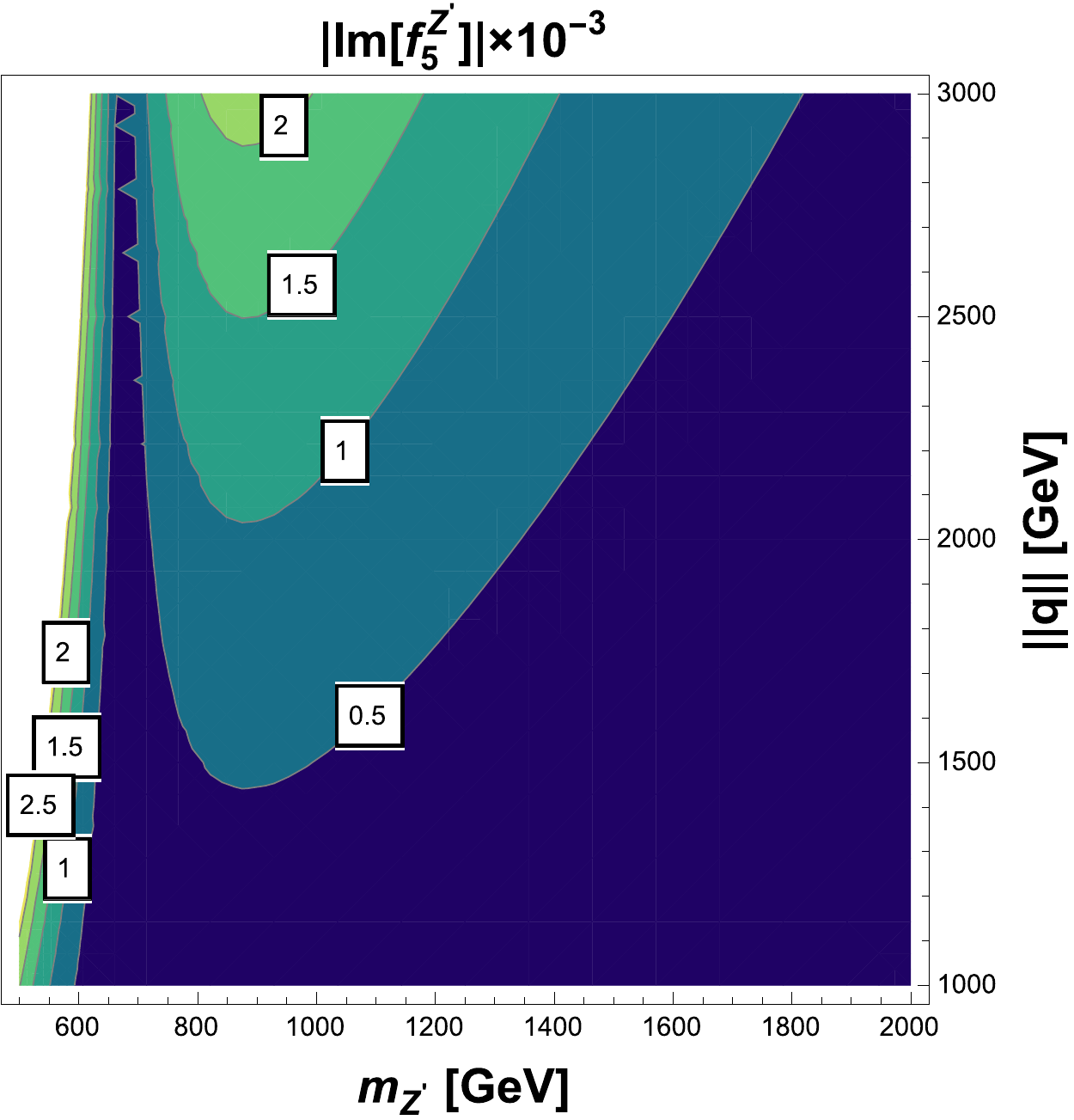}
\caption{Contour lines of the $f_5^{Z^\prime}$ form factor in the $|q|$ vs $m_{Z'}$ plane in the diagonal case. \label{plotZp2}}
\end{center}
\end{figure*}

We first analyze the behavior of the $CP$-conserving form factor $f_5^{Z'}$ in the scenario with no FCNCs (diagonal case). We show in Fig. \ref{plotZp1} the behavior of  $f_5^{Z^\prime}$ as a function of the heavy $Z'$ gauge boson transfer momentum $q^2$ (left plot) and the  $m_{Z^\prime}$ mass (right plot). We observe that the dominant contributions arise from the light quarks and leptons, whereas the top quark contribution is the smaller one as its coupling with the $Z'$ gauge boson is proportional to the  $x$ parameter, which is taken  of the order of $10^{-1}$. This behavior is also observed in the $CP$-conserving $ZZZ^\ast$ form factor in the diagonal case \cite{Gounaris:2000tb}. We also note that $f_5^{Z'}$ decreases for increasing transfer momentum $|q|$, but it increases for large values of $m_{Z^\prime}$. Since $f_5^{Z'}$ is proportional to $m_{Z^\prime}$  [see Eq. \eqref{f5ZpDiagonal}], a similar behavior is expected in the non-diagonal case. In Fig. \ref{plotZp2} we present the contour lines of the total real (left plot) and imaginary (right plot) parts of $f_5^{Z^\prime}$ in the $|q|$ vs $m_{Z^\prime}$ plane. It is observed that at high energy, both real and imaginary parts of $f_5^{Z^\prime}$ are considerably small, of the order of $10^{-2}$ and $10^{-3}$, respectively, which is true even if the mass of the heavy boson is very large. For   $m_{Z^\prime}\gg$ 3000 GeV and intermediate values of $|q|$, the value of the real part of $f_5^{Z^\prime}$ can be of the order of $O(1)$, whereas the imaginary part is of the order of $10^{-2}$.

\begin{figure*}[!hbt]
\begin{center}
\includegraphics[width=18cm]{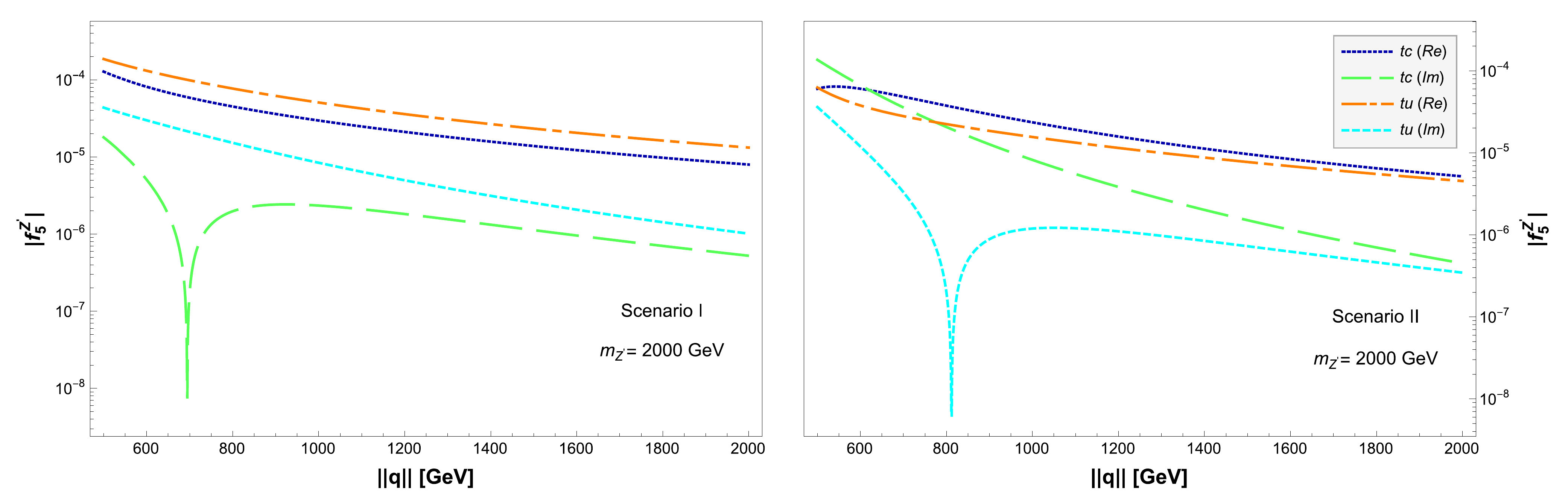}
\caption{Behavior of the $f_5^{Z^\prime}$ form factor in the non-diagonal case as a function of the transfer momentum $|q|$ of the  $Z'$ gauge boson.   \label{plotZp3}}%
\end{center}
\end{figure*}

\begin{figure*}[!hbt]
\begin{center}
\includegraphics[width=7cm]{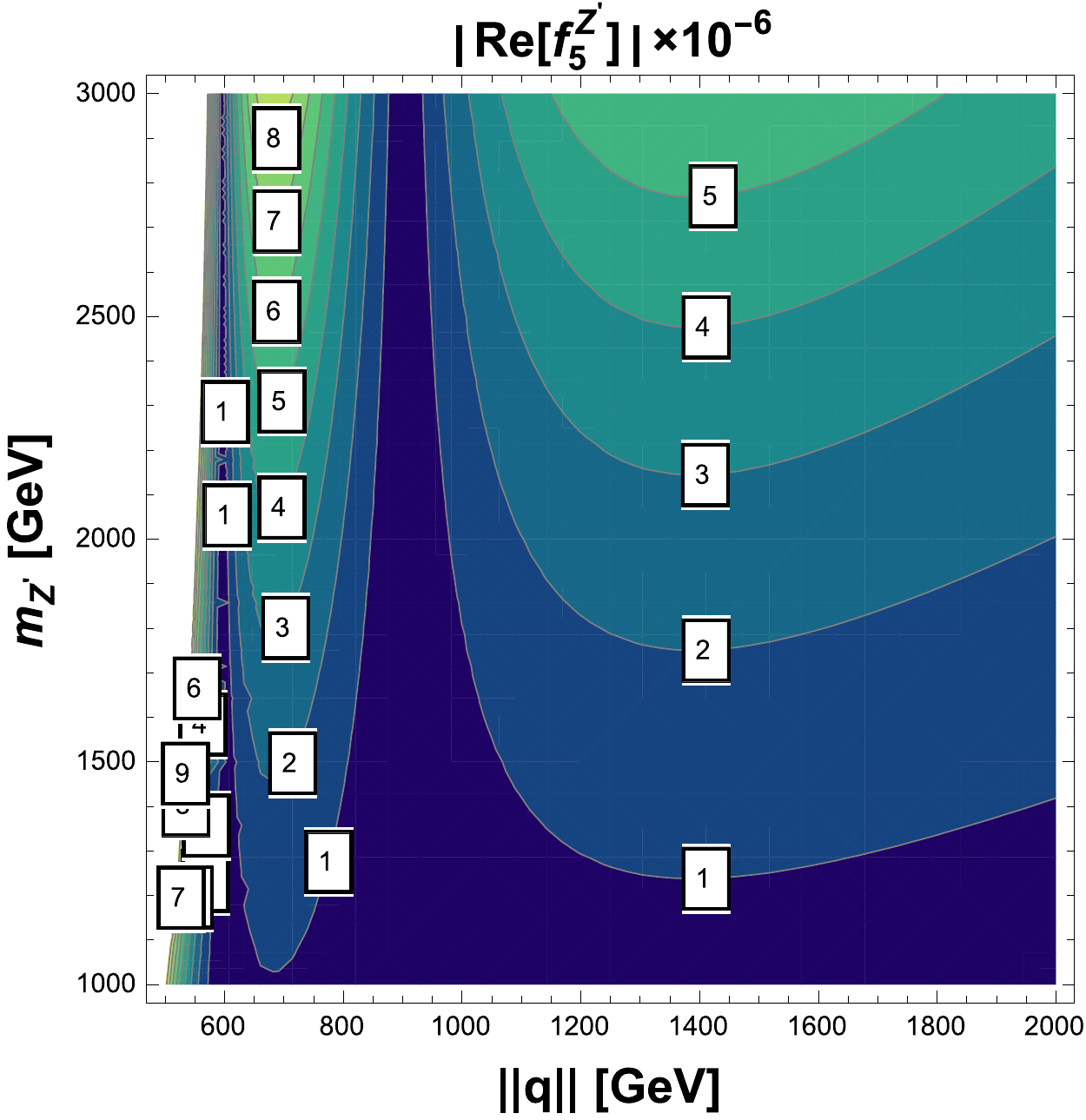}
\includegraphics[width=7cm]{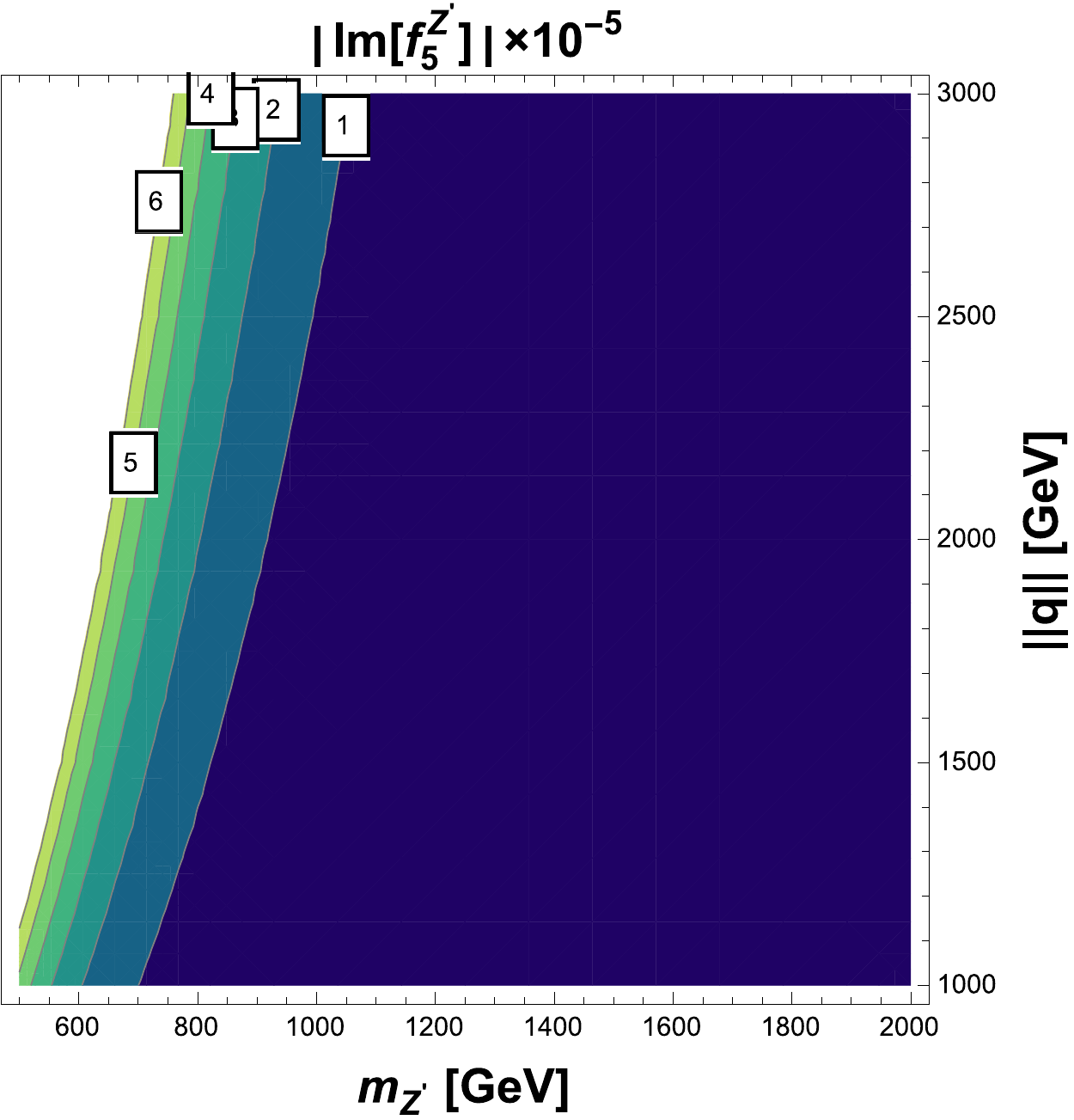}
\caption{Contour lines in the $|q|$ vs $m_{Z'}$ plane of the real and imaginary parts of the $f_5^{Z^\prime}$ form factor in the non-diagonal case and scenario II.\label{plotZp4}   }%
\end{center}
\end{figure*}

We now show in Fig. \ref{plotZp3} the form factor $f_5^{Z^\prime}$ as function of the transfer momentum $|q|$ of the $Z'$ gauge boson in the non-diagonal case. For the FCNCs couplings of the $Z$ gauge boson, we consider both scenario I (left plot) and scenario II (right plot), which were also  considered in the analysis of the $ZZZ^\ast$ vertex. As we are assuming only flavor violation in the up quark sector for the FCNCs mediated by the $Z'$ boson, we only plot this class of contributions. As in the analysis of the $ZZZ^\ast$ form factor, we only show the non-negligible contributions, which are those where the top quark runs into the loops. We observe that in both scenarios the real parts of the $Z'tc$ and $Z'tu$ contributions are of similar size, although the largest contribution is distinct in each case. We also find that in scenario I the real parts of the $Z'tc$ and $Z'tu$ contributions are of the same sign, but they are of opposite sign in scenario II. Thus, they tend to cancel each other out. As for the imaginary parts of the partial contributions, they exhibit a similar behavior in both scenarios, nevertheless there is a peak in the 600 GeV$<|q|<900$ GeV region, which is present in a  distinct contribution in each scenario. We also show in Fig \ref{plotZp4}  the contour lines in the $|q|$ vs $m_{Z^\prime}$ plane of the real (left plot) and imaginary (right plot) parts of $f_5^{Z^\prime}$ in scenario II, where it is manifest the cancellation effect between the real parts of the $Z'tc$ and $Z'tu$ contributions for $|q|$ around $900$ GeV. In this scenario the form factor $f_5^{Z'}$ can be of the order of  $10^{-6}$, though for intermediate  $|q|$ and large $m_{Z^\prime}$ it can reach values one order of magnitude larger. As for scenario I, the real and imaginary parts of $f_5^{Z'}$ are of the order $10^{-4}$ in general, but they could be larger for small energies and an ultra heavy $Z^\prime$.

\begin{figure*}[!hbt]
\begin{center}
\includegraphics[width=11cm]{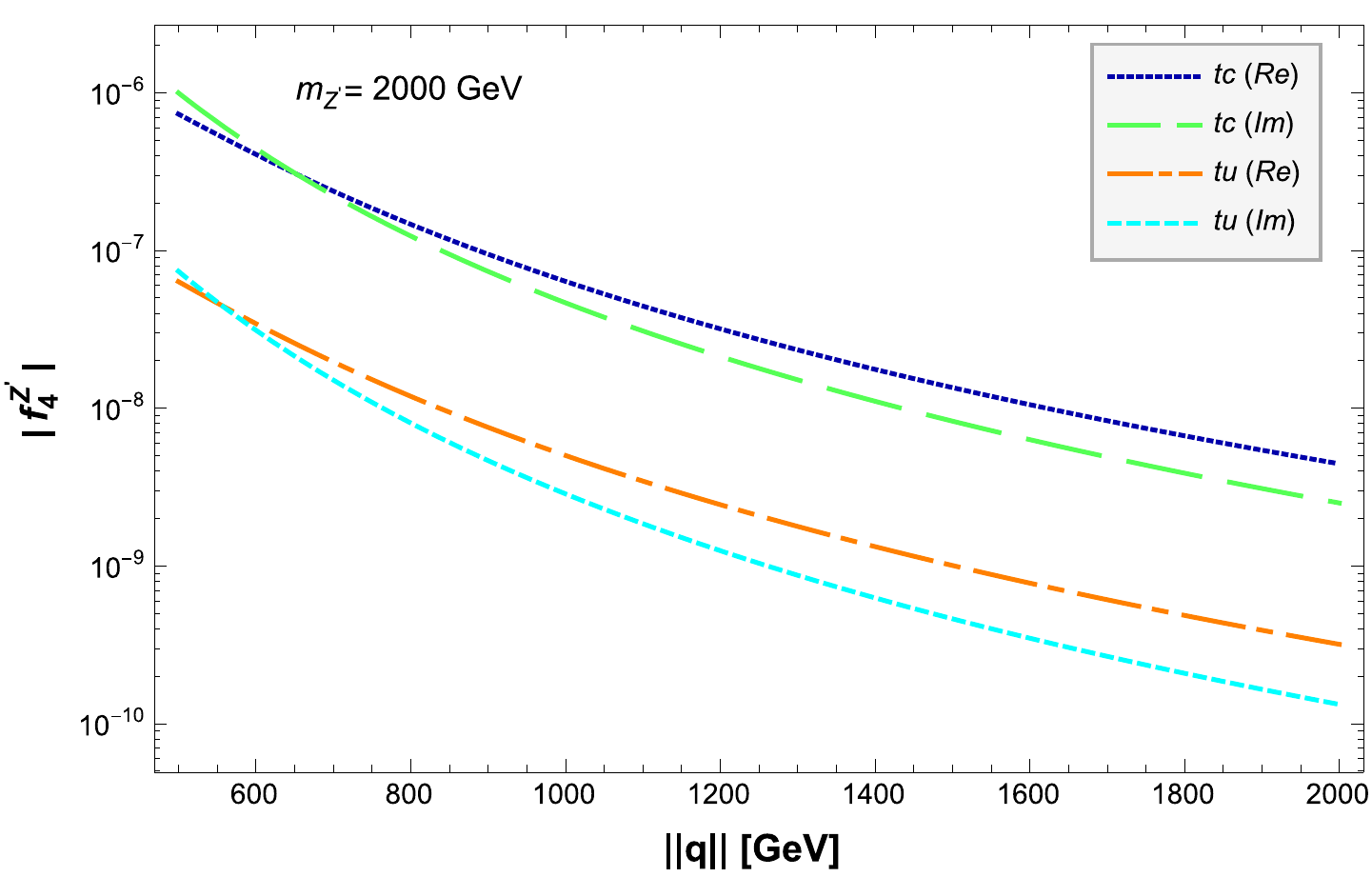}
\caption{Behavior of the $f_4^{Z^\prime}$ form factor as a function of $|q|$ in the non-diagonal case and scenario I.\label{plotZp5}   }%
\end{center}
\end{figure*}

\begin{figure*}[!hbt]
\begin{center}
\includegraphics[width=9cm]{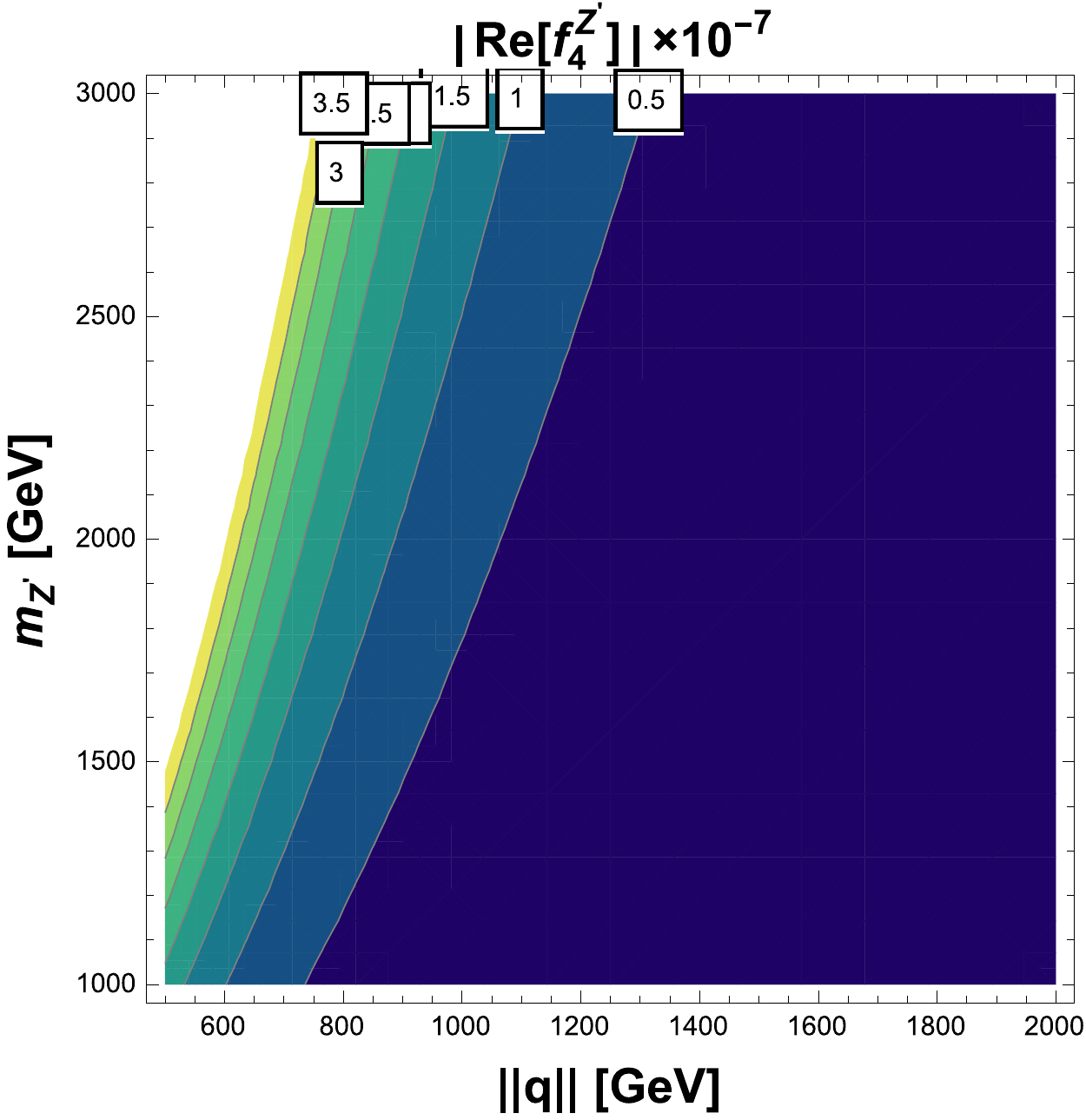}
\caption{Contour lines of the real part of $f_4^{Z^\prime}$ in the $q$ vs $m_{Z'}$ plane  in the non-diagonal case and scenario I. \label{plotZp6}  }%
\end{center}
\end{figure*}

It is also possible to induce the $CP$-violating form factor $f_4^{Z'}$ via FCNC $Z$ and $Z'$ couplings. We find that the only non-negligible contributions arise from the  $Z'tc$ and $Z'tu$ couplings, though the dominant contribution to both real and imaginary parts of $f_4^{Z'}$ is the $Z'tc$ one, which is one order of magnitude larger than the $Z'tu$ contribution.  We present in Fig. \ref{plotZp5} the form factor $f_4^{Z'}$ as a function of  $|q|$. We observe that  the real and imaginary parts behave in a  rather similar way. As was the case for the $ZZZ^*$ $CP$-violating form factor,  there is no considerable distinction between the results for scenario I and  scenario II of the FCNC $Z$ couplings, thus we only consider  scenario I in our analysis. We also show in Fig. \ref{plotZp6} the contour lines of the real  part of $f_4^{Z'}$ in the $|q|$ vs $m_{Z'}$ plane. The behavior of the imaginary part is similar as already stated. We note that at high energy  $f_4^{Z'}$ can be of the order of $10^{-7}-10^{-8}$, though it can be one order of magnitude larger at low energy and for an ultra heavy $Z'$ gauge boson. In our numerical analysis we did not extract the complex phases as in the $ZZZ^\ast$ case, since the $f_4^{Z'}$ factors depends on five distinct combinations of all of the involved phases [see Eq. \eqref{f4Zp}].

\section{Conclusions and outlook}
\label{conclusions}
We have presented a calculation of the TNGBCs $ZZV^\ast$ ($V=\gamma$, $Z$, $Z'$) in models where FCNCs couplings mediated by the $Z$ and $Z'$ gauge bosons are allowed. These TNGBCs are given in terms of one  $CP$-conserving form factor $f_5^V$ and another $CP$-violating one  $f_4^V$, for which we present analytical results in terms of Passari\-no-Velt\-man scalar functions. Such results reduce  to the contributions with diagonal $Z$ couplings already studied in the literature.  To asses the behavior of $f_4^V$ and $f_5^V$, for the numerical analysis we obtain  constraints on the FCNCs couplings of the $Z$ gauge boson to up quarks, which are the less constrained by experimental data: it is found that the current constraints on the $t\rightarrow qZ$  branching ratios obtained at the LHC translate into the following constraints on the vector and axial $Z$  couplings  $\left|g_{{VZ}}^{tu}\right|< 0.0096$ and $\left|g_{{VZ}}^{tc}\right|<0.011$. As far as  the $ZZ\gamma^\ast$ coupling  is concerned, it is found that the only non-vanishing form factor is the $CP$-conserving one $f_5^{\gamma}$, whose real and imaginary parts are of the order of $10^{-3}$, with the dominant contributions arising from the heavier up and down quarks. On the other hand, as for the $ZZZ^\ast$ coupling, both the $CP$-conserving and the $CP$ violating form factors are non-vanishing. We consider two scenarios for the FCNC $Z$ couplings (scenario I and scenario II) and  find that the magnitude of the real and imaginary parts of these form factors are of the order of  $|f_5^{Z}|\sim 10^{-6}$ and $|f_4^{Z}|\sim |{\rm Im}\left( g_{{AZ}}^{tc\ast} g_{{VZ}}^{tc} \right) |\times 10^{-5}$, with the dominant contributions arising from the non-diagonal top quark couplings. Our estimates for the FCNC contributions to the $CP$-conserving $f_5^{\gamma}$ and $f_5^{Z}$ form factors are smaller than the prediction of the SM, whereas the  $f_4^{Z}$ form factor is not induced in the SM up to  the one loop level.

We also consider the case of a new heavy neutral $Z'$ gauge boson with FCNCs and obtain the TNGBC $ZZ{Z'}^{*}$, for which we present analytical results in the case of both  diagonal  and non-diagonal $Z'$ couplings in terms of Passarino-Veltman scalar functions.  In the diagonal case we find the following numerical estimate for the  $CP$-conserving  $f_5^{Z'}$ form factor, which is the only non-vanishing,  $|{\rm Re}f_{5}^{Z'}|\sim 10^{-1}-10^{-2}$ and $|{\rm Im} f_{5}^{Z^\prime}|\sim 10^{-2}-10^{-3}$, with the dominant contributions arising from the light quarks and leptons. In the non-diagonal case we also consider two scenarios for the FCNC  couplings of the $Z$ gauge boson (scenario I and scenario II). It is found that both the real and imaginary parts of $f_{5}^{Z^\prime}$ are of the order of $10^{-4}$ in  scenario I, whereas in  scenario II  $|{\rm Re}f_{5}^{Z'}|\sim 10^{-6}$ and $|{\rm Im}f_{5}^{Z'}|\sim 10^{-5}$. In general, in the non-diagonal case  the magnitude of  both real and imaginary parts of the $f_5^{Z'}$ form factor are one order of magnitude larger for moderate energies and an ultra heavy  $Z'$ gauge boson than for high energies, with the dominant contributions arising from the $Z'tu$ and $Z'tc$ couplings. The real (imaginary) part of the non-diagonal contributions to $f_5^{Z'}$  are at least two (one) orders of magnitude smaller than the real (imaginary) parts of the diagonal contributions.

As far as  the $CP$-violating form factor $f_4^{Z'}$ is concerned, we obtain similar estimates for its real and imaginary parts,  of the order of $10^{-7}-10^{-8}$ in both scenarios of the FCNC  couplings of the $Z$ gauge boson. In closing we would like to stress that  FCNC couplings can also yield $CP$ violation  in the TNGBCs of a new neutral gauge boson, which may be on interest.

\begin{acknowledgements}
We acknowledge support from Consejo Nacional de Ciencia y Tecnolog\'ia and Sistema Nacional de Investigadores (Mexico). Partial support from Vicerrector\'ia de Investigaci\'on y Estudios de Posgrado de la Ben\'emerita Universidad Aut\'onoma de Puebla is also acknowledged.
\end{acknowledgements}

\onecolumn
\appendix

\section{Analytical form of the TNGBCs $ZZV^*$ ($V=\gamma, Z, Z'$)}
\label{AnalyticalResults}
In this appendix we present the analytical expressions for the loop functions appearing in the contributions to the  TNGBCs  $ZZV^*$ ($V=\gamma, Z, Z'$)  arising from the FCNC couplings mediated by the $Z$ gauge boson and a new heavy neutral gauge boson $Z'$. For the calculation we use the Passarino-Veltman reduction scheme.

\subsection{Passarino-Veltman results}

\subsubsection{ $ZZ\gamma^*$ coupling}
There are only contribution to the $f_5^\gamma$ form factor, which is given in Eq. \eqref{f5Photon}, where $R_{ij}$ reads
\begin{align}
\label{Rij}
    R_{ij}&=4\left(q^2-m_Z^2\right)\left(m_j^2-m_i^2 \right)\left(B_{ii}\left(q^2 \right)-B_{jj}\left(q^2 \right) \right)+2\left(q^2-2m_Z^2 \right)\left(q^2-4m_Z^2 \right)\nonumber\\
    & +2m_Z^2\left(q^2+2m_Z^2 \right) \left(B_{ii}\left(q^2\right)+B_{jj}\left(q^2\right)-2B_{ij}\left(m_Z^2\right) \right)\nonumber\\
    &+4\left(q^2-m_Z^2 \right)\left(m_i^4+m_j^4+m_Z^4-2m_i^2m_j^2-2m_i^2m_Z^2-2m_j^2m_Z^2\right)\left( C_{iji} \left(q^2 \right)+C_{jij}\left(q^2 \right)\right)\nonumber\\
    &+2q^2\left( q^2+2m_Z^2\right)\left( m_j^2 C_{iji}\left(q^2 \right)+m_i^2 C_{jij}\left(q^2 \right) \right),
\end{align}
where we have introduced the shorthand notation
\begin{align}
B_{ij}(c^2)&=B_0\left(c^2,m_i^2,m_j^2 \right),\nonumber\\
C_{ijk}\left( q^2\right)&=C_0\left(m_Z^2,m_Z^2,q^2,m_i^2,m_j^2,m_k^2 \right),
\end{align}
with $B_0$ and $C_0$ being the usual two- and three-point Passarino-Veltman scalar functions. It is useful observe the following symmetry relations
\begin{align}
B_{ij}(c^2)&=B_{ji}(c^2),\nonumber\\
C_{ijk}\left( q^2\right)&=C_{kji}\left( q^2\right),\nonumber\\
C_{iij}\left( q^2\right)&= C_{jji}\left( q^2\right),\nonumber\\
C_{iji}\left( q^2\right)&=C_{jij}\left( q^2\right).
\end{align}

In  Eq.  \eqref{Rij} it is evident that ultraviolet divergences cancel out. We have also verified that $R_{ij}$ vanishes for an on-shell photon.

\subsubsection{$ZZZ^*$ coupling}\label{Appendix2}
There are contributions to both the $CP$-conserving form factor $f_5^Z$ and the $CP$-violating one $f_4^Z$. They are given in Eqs. \eqref{f4Z} and \eqref{f5Z}, with the $R_{kij}$, and $S_{ij}$ functions given in terms of Passarino-Veltman scalar functions as follows

\begin{align}
    R_{1ij}= &1+\frac{2 \left(m_i^2-m_j^2\right) \left(2 m_Z^2+q^2\right)}{q^2
   \left(q^2-4 m_Z^2\right)}\left(B_{{ij}}\left(m_Z^2\right)-B_{{ii}}\left(m_Z^2\right)\right)\nonumber\\
    &-\frac{1}{\left(q^2-4 m_Z^2\right)
   \left(q^2-m_Z^2\right)}\Bigg[\frac{1}{q^2}C_{{iij}}\left(q^2\right) \left(-q^4
   \left(m_i^2 \left(m_i^2-m_j^2\right)-m_Z^2 \left(7 m_i^2+3 m_j^2-4
   m_Z^2\right)\right)\right.\nonumber\\
   &\left.+4 q^2 m_Z^2 \left(m_i^2 m_j^2-2 m_i^2 m_Z^2-m_j^4+m_Z^4\right)+4
   m_Z^4 \left(m_i^2-m_j^2\right)^2-2 q^6
   m_i^2\right)\nonumber\\
    &-C_{{iji}}\left(q^2\right) \left(q^2 \left(-3 m_i^2 m_j^2-3
   m_i^2 m_Z^2+m_i^4-2 m_j^2 m_Z^2+2 m_j^4+2 m_Z^4\right)\right.\nonumber\\
   &\left.+2 m_Z^2 \left(m_i^4+2 m_j^2
   m_Z^2-m_j^4-m_Z^4\right)+q^4 m_j^2\right)\nonumber\\
   &+\left(B_{{ii}}\left(q^2\right)-B_{{ii}}\left(m_Z^2\right)\right)
   \left(\left(m_i^2-m_Z^2\right) \left(2 m_Z^2+q^2\right)-2 m_j^2
   \left(q^2-m_Z^2\right)\right)\nonumber\\
   &+\left(B_{{ij}}\left(q^2\right)-B_{{ij}}\left(
   m_Z^2\right)\right) \left(\left(m_j^2-2 m_Z^2\right) \left(2 m_Z^2+q^2\right)-3 m_i^2
   \left(q^2-2 m_Z^2\right)\right)\Bigg],
\end{align}

\begin{align}
    R_{2ij}= & \frac{m_i^2}{q^2-m_Z^2} \Big[   \left(m_i^2-m_j^2-m_Z^2\right)\left(C_{{iij}}\left(q^2\right)-C_{{iji}}\left(q^2\right)\right)
   +\left(q^2-m_Z^2\right)C_{{iij}}\left(q^2\right)
     \nonumber\\
   &  +B_{{ii}}\left(q^2\right)-B_{{ii}}\left(m_Z^2\right)+2
   \left(B_{{ij}}\left(q^2\right)-B_{{ij}}\left(m_Z^2\right)\right)\Big],
\end{align}

\begin{align}
    R_{3ij}= & \frac{m_i m_j}{q^2-m_Z^2} \Big[ \left(2 m_i^2-2
   m_j^2+q^2\right)\left[C_{{iij}}\left(q^2\right)-C_{{iji}}\left(q^2\right)\right] +2\left(q^2-m_Z^2\right) C_{{iji}}\left(q^2\right)
    \nonumber\\
   & +2
   \left[B_{{ii}}\left(q^2\right)-B_{{ii}}\left(m_Z^2\right)\right]+4
   \left[B_{{ij}}\left(q^2\right)-B_{{ij}}\left(m_Z^2\right)\right]\Big],
\end{align}
and
\begin{align}
S_{ij}&=\left(4
   m_Z^2-2 q^2\right)\left[ B_{{ii}}\left(m_Z^2\right)-
   B_{{jj}}\left(m_Z^2\right)\right] \nonumber\\
   & -2 q^2 \left[B_{ii}\left(q^2\right)-B_{jj}\left(q^2\right)\right] - \left(2 m_Z^2-q^2\right) \left(2 m_i^2-2
   m_j^2+q^2\right) C_{iij}\left(q^2\right)\nonumber\\
   & -q^2  \left[q^2-2
   \left(m_i^2-m_j^2+m_Z^2\right)\right] C_{iji}\left(q^2\right)+
   \left(q^2-2 m_Z^2\right) \left(2 m_i^2-2 m_j^2-q^2\right) C_{jji}\left(q^2\right)\nonumber\\
   &+q^2
    \left[q^2-2 \left(m_j^2-m_i^2+m_Z^2\right)\right] C_{{jij}}\left(q^2\right).
\end{align}

\subsubsection{$ZZZ^{'*}$ coupling}\label{Appendix3}
The contributions to the $f_5^{Z'}$ and  $f_4^{Z'}$ form factors are given in Eqs. \eqref{f5ZpDiagonal}-\eqref{f4Zp}, with the $L_i$ $T_{kij}$ and  $U_{kij}$ functions given as follows

\begin{align}
L_{1i}&=2 m_Z^2 \big(2 \left(q^2 \left(m_i^2+m_Z^2\right)-m_Z^2
   \left(4 m_i^2+m_Z^2\right)\right)
   C_{iii}\left(q^2\right)\nonumber\\
   &-\left(2 m_Z^2+q^2\right)
   \left(B_{{ii}}\left(m_Z^2\right)-B_{{ii}}\left(q^2\right)\right)\big)-6
   q^2 m_Z^2+8 m_Z^4+q^4,
   \end{align}

\begin{align}
L_{2i}&=-4 \left(m_Z-q\right) \left(m_Z+q\right) \left(m_i^2
   \left(q^2-4 m_Z^2\right)+m_Z^4\right)
   C_{iii}\left(q^2\right)\nonumber\\
   &-2 \left(4 m_i^2 \left(q^2-4
   m_Z^2\right)+m_Z^2 \left(2 m_Z^2+q^2\right)\right)
   \left(B_{{ii}}\left(m_Z^2\right)-B_{{ii}}\left(q^2\right)\right)-6 q^2
   m_Z^2+8 m_Z^4+q^4,
   \end{align}

\begin{align}
L_{3i}&=2  \big(2 \left(m_i^2 \left(-6 q^2 m_Z^2+8
   m_Z^4+q^4\right)-2 m_Z^4 \left(m_Z^2-q^2\right) \right)
   C_{iii}\left(q^2\right)\nonumber\\
   &-2 m_Z^2 \left(2
   m_Z^2+q^2\right)
   \left(B_{{ii}}\left(m_Z^2\right)-B_{{ii}}\left(q^2\right)\right)-6 q^2
   m_Z^2+8 m_Z^4+q^4\big).
   \end{align}

\begin{align}
U_{1ij}&=-\frac{1} { q^2} \Big[
   B_{{ii}}\left(m_Z^2\right) \big(q^4 \left(2 m_i
   \left(m_i-m_j\right)+m_Z^2\right)+2 q^2 m_Z^2 \left(m_Z^2-2
   \left(m_i-m_j\right)^2\right)\nonumber\\
   &-4 m_Z^4 \left(m_i^2-m_j^2\right)\big)+q^2 \left(B_{{ij}}\left(m_Z^2\right)-2
   B_{{ij}}\left(q^2\right)\right) \big(q^2
   \left(\left(m_i-m_j\right)^2+m_Z^2\right)-4 m_Z^2 \left(m_i-m_j\right)^2\nonumber\\
   &+2
   m_Z^4\big)+C_{{iij}}\left(q^2\right) \big(q^6 m_i \left(m_j-3 m_i\right)+q^4
   \big(m_Z^2 \left(m_i(13 m_i-4 m_j)+3 m_j^2\right)-2 m_i \left(m_i-m_j\right)^2
   \left(m_i+m_j\right)\nonumber\\
   &-4 m_Z^4\big)+4 q^2 m_Z^2 \left(\left(m_i-m_j\right)^3
   \left(m_i+m_j\right)-4 m_i^2 m_Z^2+m_Z^4\right)+4 m_Z^4
   \left(m_i^2-m_j^2\right)^2\big)\nonumber\\
   &+ \left(6 q^4 m_Z^2-8 q^2
   m_Z^4-q^6\right)\Big] + (i \leftrightarrow j),
\end{align}

\begin{align}
U_{2ij}=
U_{1ij}(m_j\to -m_j),
\end{align}

\begin{align}
U_{3ij}&=2
   \Big[B_{{ii}}\left(q^2\right) \left(q^2 \left(2( m_j^2- m_j^2)+m_Z^2\right)+2
   m_Z^2 \left(m_i^2-m_j^2+m_Z^2\right)\right)-m_Z^2 \left(2 m_Z^2+q^2\right)
   B_{{ij}}\left(m_Z^2\right)\nonumber\\
   &+C_{{iji}}\left(q^2\right) \big(2 q^2
   \left(m_Z^4-m_Z^2 \left(2 m_i^2+m_j^2\right)+\left(m_i^2-m_j^2\right)^2\right)-2
   m_Z^2 \left((m_i+m_j)^2-m_Z^2\right) \left((m_i-m_j)^2-m_Z^2\right) \nonumber\\
   &+q^4 m_j^2\big)-3 q^2 m_Z^2+4 m_Z^4+\frac{1}{2}q^4\Big]+ (i \leftrightarrow j),
\end{align}

\begin{align}
U_{4ij}&=-\frac{1 }{ q^2} \Big[  m_Z^2
   B_{{ii}}\left(m_Z^2\right) \left(2 q^2 \left(2 m_i^2-2 m_j^2+m_Z^2\right)-4
   m_Z^2 \left(m_i^2-m_j^2\right) +q^4\right)\nonumber\\
   &+q^2 \left(m_Z^2 \left(2
   m_Z^2+q^2\right) \left(B_{{ij}}\left(m_Z^2\right)-2
   B_{{ij}}\left(q^2\right)\right)+ \left(6 q^2 m_Z^2-8
   m_Z^4+q^4\right)\right)\nonumber\\
   &+C_{{iij}}\left(q^2\right) \big(q^6 m_i
   \left(m_j-m_i\right)-q^4 m_Z^2 \left(8 m_i m_j-m_i^2-3 m_j^2+4 m_Z^2\right)\nonumber\\
   &-4 q^2
   m_Z^2 \left((m_i-m_j)^2-m_Z^2\right)
   \left(\left(m_i+m_j\right)^2+m_Z^2\right)+4 m_Z^4
   \left(m_i^2-m_j^2\right)^2\big)\Big]+ (i \leftrightarrow j),
\end{align}

\begin{align}
U_{5ij}&=U_{4ij}(m_j\to -m_j),
\end{align}

\begin{align}
U_{6ij}&=-2 \Big[
   B_{{ii}}\left(q^2\right) \left(-q^2
   \left(2 m_j \left(m_i+m_j\right)+m_Z^2\right)+2
   m_Z^2 \left(m_i+m_j\right) \left(3 m_i+m_j\right)-2
   m_Z^4\right)\nonumber\\
   &+B_{{ij}}\left(m_Z^2\right)
   \left(q^2
   \left(\left(m_i+m_j\right)^2+m_Z^2\right)-4 m_Z^2
   \left(m_i+m_j\right)^2+2
   m_Z^4\right)\nonumber\\
   &+C_{{iji}}\left(q^2\right)
   \big(-q^4 m_j \left(m_i+m_j\right)+2
   q^2 \left(m_Z^2 \left(3 m_i
   m_j+m_i^2+m_j^2\right)+m_j \left(m_i^2-m_j^2\right)
   \left(m_i+m_j\right)-m_Z^4\right)\nonumber\\
   &-2 m_Z^2
   \left((m_i+m_j)^2-m_Z^2\right)
   \left(\left(m_i-m_j\right) \left(3
   m_i+m_j\right)+m_Z^2\right)\big)+3 q^2 m_Z^2-4
   m_Z^4-\frac{1}{2}q^4\Big] + (i \leftrightarrow j) ,
\end{align}
and

\begin{equation}
U_{7ij}=U_{6ij}(m_j\to-m_j).
\end{equation}

\begin{align}
T_{1ij} &=   \Big[3
   \left(m_i-m_j\right) \big(q^2 \left(m_i+m_j\right)
   \left(B_{{ij}}\left(q^2\right) \left(q^2-2 m_Z^2\right)-m_Z^2
   B_{{ij}}\left(m_Z^2\right)\right)+C_{{iij}}\left(q^2\right) \big(q^4
   \big(2 m_i^2 \left(m_i+m_j\right)\nonumber\\
   &-m_Z^2 \left(3 m_i+m_j\right)\big)-2 q^2 m_Z^2
   \left(m_i+m_j\right) \left(2 m_i \left(m_i+m_j\right)-m_Z^2\right)+2 m_Z^2
   \left(m_i-m_j\right)^2 \left(m_i+m_j\right)^3+q^6
   m_i\big)\big)\nonumber\\
   &+B_{{ii}}\left(m_Z^2\right) \left(q^2 m_Z^2 \left(12 m_i m_j+7
   m_i^2-3 m_j^2-4 m_Z^2\right)-6 m_Z^2 \left(m_i^2-m_j^2\right)^2+q^4 \left(m_Z^2-4
   m_i^2\right)\right)\nonumber\\
   &-2 q^2 B_{{ii}}(0) m_i^2 \left(q^2-4 m_Z^2\right)\Big]- (i \leftrightarrow j),
\end{align}

\begin{align}
T_{2ij}&=T_{1ij}(m_j\to -m_j),
\end{align}

\begin{equation}
T_{3ij}=6 q^4 m_i m_j
   \Big[C_{{iji}}\left(q^2\right) \left(2
   \left(m_i^2-m_j^2+m_Z^2\right)-q^2\right)-2 B_{{ii}}\left(q^2\right)\Big] - (i \leftrightarrow j),
\end{equation}

\begin{align}
T_{4ij}&=  \Big[3
   \left(m_i+m_j\right) \big(C_{{iij}}\left(q^2\right) \big(q^4 \left(m_Z^2
   \left(m_i+m_j\right)+2 m_i m_j \left(m_j-m_i\right)\right)-2 q^2 m_Z^2
   \left(m_i-m_j\right) \nonumber\\
   &\times\big(2 m_i \left(m_i-m_j\right)-m_Z^2\big)+2 m_Z^2
   \left(m_i-m_j\right) \left(m_i^2-m_j^2\right)^2-q^6 m_i\big)\nonumber\\
   &-q^2
   \left(m_i-m_j\right) \big(B_{{ij}}\left(q^2\right) \left(2
   m_Z^2+q^2\right)+\left(m_Z^2-q^2\right)
   B_{{ij}}\left(m_Z^2\right)\big)\big)\nonumber\\
   &+B_{{ii}}\left(m_Z^2\right)
   \big(q^4 \left(2 m_i \left(m_i+3 m_j\right)+m_Z^2\right)-q^2 m_Z^2 \left(12 m_i
   m_j-7 m_i^2+3 m_j^2+4 m_Z^2\right)-6 m_Z^2 \left(m_i^2-m_j^2\right)^2\big)\nonumber\\
   &-2 q^2
   B_{{ii}}(0) m_i^2 \left(q^2-4 m_Z^2\right)\Big]- (i \leftrightarrow j) ,
\end{align}

\begin{align}
T_{5ij}&=T_{4ij}(m_j\to -m_j),
\end{align}

\subsection{Closed form results}
We now present the closed form of the TGNBCs presented above. We only expand the two-point scalar functions in terms  of transcendental functions as the three-point functions are too cumbersome to be expanded.
We first introduce the following auxiliary functions:

\begin{equation}
\eta\left(x\text{,} y\right)=\sqrt{x^4-4 x^2 y^2},
\end{equation}

\begin{equation}
\beta\left(x\text{,}y\text{,}z\right)=\sqrt{-2 x^2 \left(y^2+z^2\right)+x^4+\left(y^2-z^2\right){}^2},
\end{equation}

\begin{equation}
\chi\left(x\text{,} y\right)=\log \left(\frac{\sqrt{x^4-4 x^2 y^2}+2 y^2-x^2}{2 y^2}\right),
\end{equation}
and

\begin{equation}
\xi\left(x\text{,}y\text{,}z\right)=\log \left(\frac{\sqrt{-2 x^2
   \left(y^2+z^2\right)+x^4+\left(y^2-z^2\right){}^2}+x^2+y^2-z^2}{2 x
   y}\right).
\end{equation}

\subsubsection{$ZZ\gamma^\ast$ coupling}
The $R_{ij}$ function of Eq.\eqref{f5Photon} can be written as follows

\begin{align}
R_{ij}&=- \frac{1}{q^2}
   \Big[-q^2 \big(2 q^2 \left(-m_Z^2 \left(2
   m_i^2+m_j^2\right)+\left(m_i^2-m_j^2\right){}^2+m_Z^4\right)-2 m_Z^2
   \left(-m_i-m_j+m_Z\right) \left(m_i-m_j+m_Z\right)\nonumber\\
   &\times \left(-m_i+m_j+m_Z\right)
   \left(m_i+m_j+m_Z\right)+q^2 m_j^2\big)
   C_{iji}\left(q^2\right)-q^2 \big(m_i^2 \big(-2 q^2
   \left(2 m_j^2+m_Z^2\right)\nonumber\\
   &+4 m_Z^2 \left(m_j^2+m_Z^2\right)+q^4\big)+2 m_i^4
   \left(q^2-m_Z^2\right)-2 \left(m_j^2-m_Z^2\right){}^2
   \left(m_Z^2-q^2\right) \big)C_{jij}\left(q^2\right)\nonumber\\
   &+3 q^4  \log
   \left(\frac{m_i^2}{m_j^2}\right)\big(m_j^2-m_i^2\big)-\eta\left(q\text{,}m_j\right)\log \left(\chi\left(q\text{,}m_j\right)\right) \left(q^2 \left(2
   m_i^2-2 m_j^2+m_Z^2\right)+2 m_Z^2 \left(-m_i^2+m_j^2+m_Z^2\right)\right)\nonumber\\
   &+2 \eta\left(q\text{,}m_i\right)
    \log \left(\chi\left(q\text{,}m_i\right)\right) \big(m_i^2-m_j^2 \big)\big(q^2-m_Z^2 \big)+6 q^4 m_Z^2+4 \beta\left(m_i\text{,}m_j\text{,}m_Z\right) q^2 m_Z^2 \log \left(\xi\left(m_i\text{,}m_j\text{,}m_Z\right)\right)\nonumber\\
    &-\eta\left(q\text{,}m_i\right)q^2 m_Z^2
   \log \left(\chi\left(q\text{,}m_i\right)\right)-8 q^2 m_Z^4-2 \eta\left(q\text{,}m_i\right)m_Z^4 \log \left(\chi\left(q\text{,}m_i\right)\right)-q^6+2
   \beta\left(m_i\text{,}m_j\text{,}m_Z\right) q^4 \nonumber\\
   &\times\log \left(\xi\left(m_i\text{,}m_j\text{,}m_Z\right)\right)\Big].
\end{align}

\subsubsection{$ZZZ^\ast$ coupling}
The $R_{kij}$ and $S_{ij}$ functions of Eqs. \eqref{f4Z} and \eqref{f5Z} read

\begin{align}
   R_{1ij} &=\frac{1}{\left( q^2 -4  m_Z^2\right)q^2 m_Z^2 }\Bigg[ -\frac{1}{\left( q^2 -  m_Z^2\right) }\Big\{-q^2 m_Z^2 \big(q^2 \left(-3 m_i^2 \left(m_j^2+m_Z^2\right)+m_i^4+2 \left(-m_j^2
   m_Z^2+m_j^4+m_Z^4\right)\right)\nonumber\\
   &-2 m_Z^2
   \left(\left(m_j^2-m_Z^2\right){}^2-m_i^4\right)+q^4 m_j^2\big)
   C_{iji}\left(q^2\right)+m_Z^2 \big(q^4 \big(m_i^2
   \left(m_j^2+7 m_Z^2\right)-m_i^4+3 m_j^2 m_Z^2-4 m_Z^4\big)\nonumber\\
   &+4 q^2 m_Z^2 \left(m_i^2
   \left(m_j^2-2 m_Z^2\right)-m_j^4+m_Z^4\right)+4 m_Z^4 \left(m_i^2-m_j^2\right){}^2-2
   q^6 m_i^2\big) C_{iij}\left(q^2\right)-m_Z^2q^2\big(q^4-5q^2m_Z^2+m_Z^4 \big)\nonumber \\
   &+ \eta\left(q\text{,} m_i\right)  m_Z^2 \left(q^2 \left(m_i^2-2 m_j^2-m_Z^2\right)+2 m_Z^2
   \left(m_i^2+m_j^2-m_Z^2\right)\right)\log \left(\chi\left(q\text{,} m_i\right)\right)\nonumber\\
   &+\eta\left(m_Z\text{,} m_i\right)  \left(m_i^2 \left(q^4-4 m_Z^4\right)+m_Z^2 \left(4 m_j^2
   \left(m_Z^2-q^2\right)+2 q^2 m_Z^2+q^4\right)\right)\log \left(\chi\left(m_Z\text{,} m_i\right)\right)\nonumber\\
   &+\beta\left(m_i\text{,}m_j\text{,}m_Z\right)  \left(q^4 \left(m_i^2+m_j^2+2 m_Z^2\right)+4 m_Z^4 \left(m_i^2-m^2_j\right)+4 q^2 m_Z^2 \left(m_Z^2-2
   m_i^2\right)\right) \log \left(\xi\left(m_i\text{,}m_j\text{,}m_Z\right)\right)\nonumber\\
   &+\beta\left(m_i\text{,}m_j\text{,}q\right) m_Z^2 \left(m_i^2 \left(6 m_Z^2-3 q^2\right)+\left(m_j^2-2 m_Z^2\right)
   \left(2 m_Z^2+q^2\right)\right)\log \left(\xi\left(m_i\text{,}m_j\text{,}q\right)\right)\Big\}\nonumber \\
    & - \frac{\left(m_i^2-m^2_j\right) }{2}
   \Big\{m_i^2 \left(10 m_Z^2-q^2\right)-\left(m_j^2+4 m_Z^2\right) \left(2
   m_Z^2+q^2\right)\Big\}\log \left(\frac{m_i^2}{m_j^2}\right)\Bigg],
\end{align}

\begin{align}
 R_{2ij}   &= \frac{m_i^2}{q^2 m_Z^2}\Bigg[-\frac{1}{
   \left(m_Z^2-q^2\right) } \Big\{q^2 m_Z^2 \left(m_i^2-m_j^2-2 m_Z^2+q^2\right)
   C_{iij}\left(q^2\right)+q^2 m_Z^2
   \left(-m_i^2+m_j^2+m_Z^2\right)
   C_{iji}\left(q^2\right)\nonumber\\
   &+m_Z^2 \big(2 \beta\left(m_i\text{,}m_j\text{,}q\right) \log
   \left(\xi\left(m_i\text{,}m_j\text{,}q\right)\right) +\eta\left(q\text{,} m_i\right) \log \left(\chi\left(q\text{,} m_i\right)\right)\big)\nonumber\\
   &-q^2 \left(2 \beta\left(m_i\text{,}m_j\text{,}m_Z\right)
   \log \left(\xi\left(m_i\text{,}m_j\text{,}m_Z\right)\right)+\eta\left( m_Z\text{,} m_i\right) \log \left(\chi\left(m_Z\text{,} m_i\right)\right)\right)\Big\}   + \left(m_i^2-m_j^2\right)  \log
   \left(\frac{m_i^2}{m_j^2}\right)\Bigg],
\end{align}

\begin{align}
   R_{3ij} &=\frac{m_i m_j}{q^2 m_Z^2}\Bigg[ -\frac{1}{ \left(m_Z^2-q^2\right) } \Big\{q^2 m_Z^2 \left(2 m_i^2-2 m_j^2+q^2\right)
   C_{iij}\left(q^2\right)\nonumber \\
   &+q^2 m_Z^2 \left(q^2-2
   \left(m_i^2-m_j^2+m_Z^2\right)\right)
   C_{iji}\left(q^2\right)+2 m_Z^2 \big(2 \beta\left(m_i\text{,}m_j\text{,}q\right)
   \log \left(\xi\left(m_i\text{,}m_j\text{,}q\right)\right) \nonumber\\
   &+\eta\left(q\text{,} m_i\right) \log \left(\chi\left(q\text{,} m_i\right)\right)\big)-2 q^2 \big(2 \beta
   _1 \log \left(\xi\left(m_i\text{,}m_j\text{,}m_Z\right)\right)+\eta\left( m_Z\text{,} m_i\right) \log \left(\chi\left(m_Z\text{,} m_i\right)\right)\big)\Big\} \nonumber  \\
    &  +2  \left(m_i^2-m_j^2\right)  \log
   \left(\frac{m_i^2}{m_j^2}\right)\Bigg],
\end{align}

and

\begin{align}
S_{ij}&=-C_{iij}\left(q^2\right) \left(2 m_Z^2-q^2\right) \left(2 m_i^2-2
   m_j^2+q^2\right)+q^2 C_{iji}\left(q^2\right) \left(-\left(q^2-2
   \left(m_i^2-m_j^2+m_Z^2\right)\right)\right)\nonumber\\
   &+C_{ijj}\left(q^2\right)
   \left(q^2-2 m_Z^2\right) \left(2 m_i^2-2 m_j^2-q^2\right)+q^2
   C_{jij}\left(q^2\right) \left(2 m_i^2-2 \left(m_j^2+m_Z^2\right)+q^2\right)\nonumber\\
   &-2 \eta\left(q\text{,} m_i\right) \log \left(\chi\left(q\text{,} m_i\right)\right)+2 \eta\left(q\text{,} m_j\right) \log \left(\chi\left(q\text{,} m_j\right)\right)+4
   \left(q^2-m_Z^2\right) \log \left(\frac{m_i^2}{m_j^2}\right)\nonumber\\
   &+\frac{2\left(2
   m_Z^2-q^2\right)}{m_Z^2} \Big\{\eta\left(m_Z\text{,} m_i\right)
   \log \left(\chi\left(m_Z\text{,} m_i\right)\right)-\eta\left(m_Z\text{,} m_j\right) \log \left(\chi\left(m_Z\text{,} m_j\right)\right)\Big\}.
\end{align}

\subsubsection{$ZZZ'^\ast$ coupling}
Finally, the $L_i$ $T_{kij}$ and  $U_{kij}$ functions of  Eqs. \eqref{f5ZpDiagonal}-\eqref{f4Zp}  are given as follows

\begin{align}
L_{1i}&=\frac{1}{q^2}\Big\{4 q^2 m_Z^2 C_{iii}\left(q^2\right) \left(q^2
   \left(m_i^2+m_Z^2\right)-m_Z^2 \left(4 m_i^2+m_Z^2\right)\right)-6 q^4 m_Z^2+2
   \left(2 m_Z^2+q^2\right) \big(\eta\left(q\text{,} m_i\right) m_Z^2 \log \left(\chi\left(q\text{,} m_i\right)\right)\nonumber\\
   &-\eta\left(m_Z\text{,} m_i\right) q^2
   \log \left(\chi\left(m_Z\text{,} m_i\right)\right)\big)+8 q^2 m_Z^4+q^6\Big\},
\end{align}

\begin{align}
    L_{2i}&=\frac{1}{q^2 m_Z^2}\Big\{-4 q^2 m_Z^2 C_{iii}\left(q^2\right) \left(m_Z^2-q^2\right)
   \left(m_i^2 \left(q^2-4 m_Z^2\right)+m_Z^4\right)+2 \big(4 m_i^2 \left(q^2-4
   m_Z^2\right)\nonumber\\
   &+m_Z^2 \left(2 m_Z^2+q^2\right)\big) \left(\eta\left(q\text{,} m_i\right) m_Z^2 \log
   \left(\chi\left(q\text{,} m_i\right)\right)-\eta\left(m_Z\text{,} m_i\right) q^2 \log \left(\chi\left(m_Z\text{,} m_i\right)\right)\right)+q^6 m_Z^2\nonumber\\
   &-6 q^4
   m_Z^4+8 q^2 m_Z^6   \Big\},
\end{align}

\begin{align}
    L_{3i}&=\frac{2}{q^2} \Big\{2 q^2 C_{iii}\left(q^2\right) \left(m_i^2 \left(-6 q^2 m_Z^2+8
   m_Z^4+q^4\right)-2 m_Z^4 \left(m^2_Z-q^2\right) \right)-6 q^4 m_Z^2\nonumber\\
   &+2
   \left(2 m_Z^2+q^2\right) \left(\eta\left(q\text{,} m_i\right) m_Z^2 \log \left(\chi\left(q\text{,} m_i\right)\right)-\eta\left(m_Z\text{,} m_i\right) q^2
   \log \left(\chi\left(m_Z\text{,} m_i\right)\right)\right)+8 q^2 m_Z^4+q^6\Big\},
\end{align}

\begin{align}
  U_{1ij}  &=-\frac{1}{q^2}\Big\{ C_{iij}\left(q^2\right) \big(q^6 m_i \left(m_j-3 m_i\right)+q^4
   \left(m_Z^2 \left(-4 m_i m_j+13 m_i^2+3 m_j^2\right)-2 m_i \left(m_i-m_j\right){}^2
   \left(m_i+m_j\right)-4 m_Z^4\right)\nonumber\\
   &+4 q^2 m_Z^2 \left(\left(m_i-m_j\right){}^3
   \left(m_i+m_j\right)-4 m_i^2 m_Z^2+m_Z^4\right)+4 m_Z^4
   \left(m_i^2-m_j^2\right){}^2\big)+C_{jji}\left(q^2\right) \big(q^6 m_j
   \left(m_i-3 m_j\right)\nonumber\\
   &+q^4 \left(m_Z^2 \left(-4 m_i m_j+3 m_i^2+13 m_j^2\right)-2 m_j
   \left(m_i-m_j\right){}^2 \left(m_i+m_j\right)-4 m_Z^4\right)+4 q^2 m_Z^2
   \big(-\left(m_i-m_j\right){}^3 \left(m_i+m_j\right)\nonumber\\
   &-4 m_j^2 m_Z^2+m_Z^4\big)+4
   m_Z^4 \left(m_i^2-m_j^2\right){}^2\big)-2 \left(-6 q^4 m_Z^2+8 q^2
   m_Z^4+q^6\right)\Big\}-\frac{2}{m_Z^2}\Big\{ \beta\left(m_i\text{,}m_j\text{,}m_Z\right) \nonumber\\
   &\times\log \left(\xi\left(m_i\text{,}m_j\text{,}m_Z\right)\right) \left(q^2
   \left(\left(m_i-m_j\right){}^2+m_Z^2\right)-4 m_Z^2 \left(m_i-m_j\right){}^2+2
   m_Z^4\right)\Big\}+\frac{4 }{q^2}\Big\{\beta\left(m_i\text{,}m_j\text{,}q\right)\nonumber\\
   &\times \log \left(\xi\left(m_i\text{,}m_j\text{,}q\right)\right) \left(q^2
   \left(\left(m_i-m_j\right){}^2+m_Z^2\right)-4 m_Z^2 \left(m_i-m_j\right){}^2+2
   m_Z^4\right)\Big\}\nonumber\\
   &+\frac{\left(m_i^2-m_j^2\right) \left(2
   m_Z^2-q^2\right)}{q^2 m_Z^2} \log \left(\frac{m_i^2}{m_j^2}\right) \Big\{q^2
   \left(-\left(\left(m_i-m_j\right){}^2+2 m_Z^2\right)\right)-4 m_Z^2
   \left(m_Z^2-(m_i-m_j)^2\right) \Big\}\nonumber\\
   &+\frac{\eta\left(m_Z\text{,} m_i\right)
   \log \left(\chi\left(m_Z\text{,} m_i\right)\right)}{q^2 m_Z^2} \Big\{q^4 \left(-\left(2 m_i
   \left(m_i-m_j\right)+m_Z^2\right)\right)-2 q^2 m_Z^2 \left(m_Z^2-2
   \left(m_i-m_j\right){}^2\right)\nonumber\\
   &+4 m_Z^4 \left(m_i^2-m^2_j\right)
  \Big\}-\frac{\eta\left(m_Z\text{,} m_j\right) \log \left(\chi\left(m_Z\text{,} m_j\right)\right)}{q^2 m_Z^2}
   \Big\{q^4 \left(2 m_j \left(m_j-m_i\right)+m_Z^2\right)\nonumber\\
   &+2 q^2 m_Z^2 \left(m_Z^2-2
   \left(m_i-m_j\right){}^2\right)+4 m_Z^4 \left(m_i^2-m^2_j\right)
   \Big\},
\end{align}

\begin{align}
  U_{2ij}  &=U_{1ij}(m_j\to -m_j),
\end{align}

\begin{align}
  U_{3ij}  &=2 \Big\{(C_{iji}\left(q^2\right) \big(2 q^2 \left(-m_Z^2 \left(2
   m_i^2+m_j^2\right)+\left(m_i^2-m_j^2\right){}^2+m_Z^4\right)-2 m_Z^2
    \left(-(m_i-m_j)^2+m_Z^2\right)  \nonumber\\
    &\times
  \left(-(m_i+m_j)^2+m_Z^2\right)+q^4 m_j^2\big)+C_{jij}\left(q^2\right) \big(m_i^2
   \left(-2 q^2 \left(2 m_j^2+m_Z^2\right)+4 m_Z^2 \left(m_j^2+m_Z^2\right)+q^4\right)\\
   &+2
   m_i^4 \left(q^2-m_Z^2\right)-2 \left(m_j^2-m^2_Z\right){}^2
   \left(m_Z^2-q^2\right) \big)-6 q^2 m_Z^2+8 m_Z^4+q^4\Big\}\nonumber\\
   &+\frac{2
   \eta\left(q\text{,} m_i\right) \log \left(\chi\left(q, m_i\right)\right)}{q^2} \Big\{q^2 \left(-2 m_i^2+2 m_j^2+m_Z^2\right)+2
   m_Z^2 \left(m_i^2-m_j^2+m_Z^2\right)\Big\}\nonumber\\
   &+\frac{2 \eta\left(q, m_j\right) \log \left(\chi\left(q, m_j\right)\right)}{q^2} \Big\{q^2 \left(2 m_i^2-2 m_j^2+m_Z^2\right)+2 m_Z^2
   \left(-m_i^2+m_j^2+m_Z^2\right)\Big\}\nonumber\\
   &+6 q^2 \left(m^2_i-m^2_j\right)
    \log \left(\frac{m_i^2}{m_j^2}\right)-4 \beta\left(m_i,m_j,m_Z\right) \log \left(\xi\left(m_i,m_j,m_Z\right)\right) \left(2 m_Z^2+q^2\right),
\end{align}

\begin{align}
  U_{4ij}  &=-\frac{1}{q^2}\Big\{C_{iij}\left(q^2\right) \big(q^6 m_i \left(m_j-m_i\right)+q^4 m_Z^2
   \left(-8 m_i m_j+m_i^2+3 m_j^2-4 m_Z^2\right)+4 q^2 m_Z^2 \left(-(m_i-m_j)^2+m^2_Z\right)\nonumber\\
   &\times \left(\left(m_i+m_j\right){}^2+m_Z^2\right)+4 m_Z^4
   \left(m_i^2-m_j^2\right){}^2\big)+C_{jji}\left(q^2\right) \big(q^6 m_j
   \left(m_i-m_j\right)+q^4 m_Z^2 \big(-8 m_i m_j+3 m_i^2+m_j^2\nonumber\\
   &-4 m_Z^2\big)+4 q^2
   m_Z^2  \left(-(m_i-m_j)^2+m^2_Z\right)
   \left(\left(m_i+m_j\right){}^2+m_Z^2\right)+4 m_Z^4
   \left(m_i^2-m_j^2\right){}^2\big)-2 \big(-6 q^4 m_Z^2+8 q^2
   m_Z^4\nonumber\\
   &+q^6\big)\Big\}-\frac{\eta\left(m_Z, m_i\right) \log \left(\chi\left(m_Z, m_i\right)\right)}{q^2} \Big\{2 q^2 \left(2
   m_i^2-2 m_j^2+m_Z^2\right)-4 m_Z^2 \left(m_i^2-m^2_j\right)+q^4\Big\}\nonumber\\
   &-\frac{\eta\left(m_Z, m_j\right) \log \left(\chi\left(m_Z, m_j\right)\right) }{q^2}\Big\{2
   q^2 \left(-2 m_i^2+2 m_j^2+m_Z^2\right)+4 m_Z^2 \left(m_i^2-m^2_j\right)+q^4\Big\}\nonumber\\
   &+\frac{\left(m_i^2-m^2_j\right)
   \left(4 q^2 m_Z^2-8 m_Z^4+q^4\right)}{q^2} \log \left(\frac{m_i^2}{m_j^2}\right)-2
   \beta\left(m_i,m_j,m_Z\right) \log \left(\xi\left(m_i,m_j,m_Z\right)\right) \left(2 m_Z^2+q^2\right)\nonumber\\
   &+\log \left(\xi\left(m_i,m_j,q\right)\right)
   \Big\{\frac{8 \beta\left(m_i,m_j,q\right) m_Z^4}{q^2}+4 \beta\left(m_i,m_j,q\right) m_Z^2\Big\},
\end{align}

\begin{align}
  U_{5ij}  &=U_{4ij}(m_j\to -m_j),
\end{align}

\begin{align}
  U_{6ij}  &=2 \Big\{C_{iji}\left(q^2\right) \big(q^4 m_j \left(m_i+m_j\right)-2 q^2
   \left(m_Z^2 \left(3 m_i m_j+m_i^2+m_j^2\right)+m_j \left(m_i-m_j\right)
   \left(m_i+m_j\right){}^2-m_Z^4\right)\nonumber\\
   &-2 m_Z^2
   \left(-(m_i+m_j)^2+m^2_Z\right) \left(\left(m_i-m_j\right) \left(3
   m_i+m_j\right)+m_Z^2\right)\big)+C_{jij}\left(q^2\right) \big(q^4 m_i
   \left(m_i+m_j\right)\nonumber\\
   &+2 q^2 \left(-m_Z^2 \left(3 m_i m_j+m_i^2+m_j^2\right)+m_i
   \left(m_i-m_j\right) \left(m_i+m_j\right){}^2+m_Z^4\right)-2 m_Z^2
    \left(-(m_i+m_j)^2+m^2_Z\right) \nonumber\\
    &\times\left(m_Z^2-\left(m_i-m_j\right)
   \left(m_i+3 m_j\right)\right)\big)-6 q^2 m_Z^2+8 m_Z^4+q^4\Big\}
   \nonumber\\
   &-\frac{4 \beta\left(m_i,m_j,m_Z\right)
   \log \left(\xi\left(m_i,m_j,m_Z\right)\right)}{m_Z^2} \Big\{q^2 \left(\left(m_i+m_j\right){}^2+m_Z^2\right)-4
   m_Z^2 \left(m_i+m_j\right){}^2+2 m_Z^4\Big\}\nonumber\\
   &+\frac{2 \eta\left(q, m_i\right) \log \left(\chi\left(q, m_i\right)\right) \left(q^2 \left(2 m_j \left(m_i+m_j\right)+m_Z^2\right)-2 m_Z^2
   \left(m_i+m_j\right) \left(3 m_i+m_j\right)+2 m_Z^4\right)}{q^2}\nonumber\\
   &+\frac{2 \eta\left(q, m_j\right) \log
   \left(\chi\left(q, m_j\right)\right) \left(q^2 \left(2 m_i \left(m_i+m_j\right)+m_Z^2\right)-2 m_Z^2
   \left(m_i+m_j\right) \left(m_i+3 m_j\right)+2 m_Z^4\right)}{q^2}\nonumber\\
   &-\frac{2
   \left(m_i^2-m^2_j\right) }{m_Z^2} \log \left(\frac{m_i^2}{m_j^2}\right)
   \Big\{4 m_Z^2 \left((m_i+m_j)^2-m^2_Z\right) -q^2
   \left(\left(m_i+m_j\right){}^2+2 m_Z^2\right)\Big\},
\end{align}

\begin{align}
\label{}
  U_{7ij}  &=U_{6ij}(m_j\to -m_j),
\end{align}

\begin{align}
    T_{1ij}& =\left(m_i-m_j\right) \Big\{3 C_{iij}\left(q^2\right) \big(q^4 \left(2 m_i^2
   \left(m_i+m_j\right)-m_Z^2 \left(3 m_i+m_j\right)\right)-2 q^2 m_Z^2
   \left(m_i+m_j\right) \left(2 m_i \left(m_i+m_j\right)-m_Z^2\right)\nonumber\\
   &+2 m_Z^2
   \left(m_i-m_j\right){}^2 \left(m_i+m_j\right){}^3+q^6 m_i\big)+3
   C_{jji}\left(q^2\right) \big(q^4 \left(2 m_j^2 \left(m_i+m_j\right)-m_Z^2
   \left(m_i+3 m_j\right)\right)\nonumber\\
   &-2 q^2 m_Z^2 \left(m_i+m_j\right) \left(2 m_j
   \left(m_i+m_j\right)-m_Z^2\right)+2 m_Z^2 \left(m_i-m_j\right){}^2
   \left(m_i+m_j\right){}^3+q^6 m_j\big)\nonumber\\
   &+4 q^2 \left(m_i+m_j\right) \left(q^2-4
   m_Z^2\right)\Big\}+6 \beta\left(m_i,m_j,m_Z\right) q^2 \log \left(\xi\left(m_i,m_j,m_Z\right)\right)
   \left(m_j^2-m_i^2\right)\nonumber\\
   &-6 \beta\left(m_i,m_j,q\right) \log \left(\xi\left(m_i,m_j,q\right)\right) \left(m_i^2-m^2_j\right)
    \left(2 m_Z^2-q^2\right)\nonumber\\
    &+\frac{\eta\left(m_Z, m_i\right) \log \left(\chi\left(m_Z, m_i\right)\right)
   }{m_Z^2}\Big\{q^2 m_Z^2 \left(12 m_i m_j+7 m_i^2-3 m_j^2-4 m_Z^2\right)-6 m_Z^2
   \left(m_i^2-m_j^2\right){}^2+q^4 \left(m_Z^2-4 m_i^2\right)\Big\}\nonumber\\
   &+\frac{\eta
   \left(m_Z, m_j\right) \log \left(\chi\left(m_Z, m_j\right)\right) }{m_Z^2}\Big\{q^2 m_Z^2 \left(-12 m_i m_j+3 m_i^2-7 m_j^2+4
   m_Z^2\right)+6 m_Z^2 \left(m_i^2-m_j^2\right){}^2+q^4 \left(4
   m_j^2-m_Z^2\right)\Big\}\nonumber\\
   &+\log \left(\frac{m_i^2}{m_j^2}\right) \Big\{q^4
   \left(3 \left(m_i^2+m_j^2\right)-m_Z^2\right)-2 q^2 m_Z^2 \left(3
   \left(m_i+m_j\right){}^2-2 m_Z^2\right)+12 m_Z^2 \left(m_i^2-m_j^2\right){}^2\Big\},
\end{align}

\begin{align}
\label{}
    T_{2ij}&= T_{1ij}(m_j\to -m_j),
\end{align}

\begin{align}
    T_{3ij}&= 6 q^4 m_i m_j \Big\{C_{iji}\left(q^2\right) \left(2
   \left(m_i^2-m_j^2+m_Z^2\right)-q^2\right)+C_{jij}\left(q^2\right) \left(2
   m_i^2-2 \left(m_j^2+m_Z^2\right)+q^2\right)\Big\}\nonumber\\
   &+12 q^4 m_i m_j \log
   \left(\frac{m_i^2}{m_j^2}\right)-12 \eta\left(q, m_i\right) q^2 m_i m_j \log \left(\chi\left(q, m_i\right)\right)+12
   \eta\left(q, m_j\right) q^2 m_i m_j \log \left(\chi\left(q, m_j\right)\right),
\end{align}

\begin{align}
    T_{4ij}&= \left(m_i+m_j\right) \Big\{3 C_{iij}\left(q^2\right) \big(q^4 \left(m_Z^2
   \left(m_i+m_j\right)+2 m_i m_j \left(m_j-m_i\right)\right)-2 q^2 m_Z^2
   \left(m_i-m_j\right) \left(2 m_i \left(m_i-m_j\right)-m_Z^2\right)\nonumber\\
   &+2 m_Z^2
   \left(m_i-m_j\right){}^3 \left(m_i+m_j\right){}^2+q^6 \left(-m_i\right)\big)+3
   C_{jji}\left(q^2\right) \big(-q^4 \left(m_Z^2 \left(m_i+m_j\right)+2 m_i m_j
   \left(m_i-m_j\right)\right)\nonumber\\
   &+2 q^2 m_Z^2 \left(m_i-m_j\right) \left(2 m_j
   \left(m_i-m_j\right)+m_Z^2\right)+2 m_Z^2 \left(m_i-m_j\right){}^3
   \left(m_i+m_j\right){}^2+q^6 m_j\big)\nonumber\\
   &+4 q^2 \left(m_i-m_j\right) \left(q^2-4
   m_Z^2\right)\Big\}+\frac{6 \beta\left(m_i,m_j,m_Z\right) q^2 \log \left(\xi\left(m_i,m_j,m_Z\right)\right)}{m_Z^2}
   \left(m_i^2-m^2_j\right)  \left(q^2-m^2_Z\right)\nonumber\\
   &-6 \beta\left(m_i,m_j,q\right) \log \left(\xi\left(m_i,m_j,q\right)\right) \left(m_i^2-m^2_j\right)
    \left(2 m_Z^2+q^2\right)+\frac{\eta\left(m_Z, m_i\right) \log \left(\chi\left(m_Z, m_i\right)\right)}{m_Z^2}
   \Big\{q^4 \big(2 m_i \left(m_i+3 m_j\right)+m_Z^2\big)\nonumber\\
   &-q^2 m_Z^2 \left(12 m_i
   m_j-7 m_i^2+3 m_j^2+4 m_Z^2\right)-6 m_Z^2
   \left(m_i^2-m_j^2\right){}^2\Big\}+\frac{\eta\left(m_Z, m_j\right) \log \left(\chi\left(m_Z, m_j\right)\right)
   }{m_Z^2}\Big\{q^4 \big(-\big(2 m_j \nonumber\\
   &\times\left(3 m_i+m_j\right)+m_Z^2\big)\big)+q^2 m_Z^2
   \left(12 m_i m_j+3 m_i^2-7 m_j^2+4 m_Z^2\right)+6 m_Z^2
   \left(m_i^2-m_j^2\right){}^2\Big\}\nonumber\\
   &+\frac{1}{m_Z^2}\log
   \left(\frac{m_i^2}{m_j^2}\right) \Big\{q^4 \left(-\left(6 m_i m_j m_Z^2+3
   \left(m_i^2-m_j^2\right){}^2+m_Z^4\right)\right)+2 q^2 m_Z^2 \big(-3 m_Z^2
   \left(m_i-m_j\right){}^2+3 \left(m_i^2-m_j^2\right){}^2\nonumber\\
   &+2 m_Z^4\big)+12 m_Z^4
   \left(m_i^2-m_j^2\right){}^2\Big\},
\end{align}
and
\begin{align}
    T_{5ij}&= T_{4ij}(m_j\to -m_j),
\end{align}

\bibliographystyle{spphys}
\bibliography{biblio1}

\end{document}